# First-principles calculations to investigate structural, elastic, electronic, thermodynamic, and thermoelectric properties of CaPd$_3$B$_4$O$_{12}$ (B = Ti, V) perovskite

M.H.K. Rubel[1,*], M.A. Hossain[2], M. Khalid Hossain[3,**], K.M. Hossain[1], A.A. Khatun[1], M.M. Rahaman[1], Md. Ferdous Rahman[4], M.M. Hossain[5], J. Hossain[6]

[1]*Department of Materials Science and Engineering, University of Rajshahi, Rajshahi 6205, Bangladesh*
[2]*Department of Physics, Mawlana Bhashani Science and Technology University, Tangail 1902, Bangladesh*
[3]*Atomic Energy Research Establishment, Bangladesh Atomic Energy Commission, Dhaka 1349, Bangladesh*
[4]*Department of Electrical and Electronic Engineering, Begum Rokeya University, Rangpur 5400, Bangladesh*
[5]*Department of Physics, Chittagong University of Engineering and Technology, Chittagong 4349, Bangladesh*
[6]*Department of Electrical and Electronic Engineering, University of Rajshahi, Rajshahi 6205, Bangladesh*

Correspondence: * mhk_mse@ru.ac.bd (*M.H.K. Rubel*), and **khalid.baec@gmail.com (*M.K. Hossain*)

**Research highlights**

- A systematic *ab*-initio study has been carried out on cubic CaPd$_3$B$_4$O$_{12}$ (B = Ti, V) quadruple perovskite via the DFT method
- Metallic and semiconducting characteristics have been observed for CPVO and CPTO materials, respectively
- CaPd$_3$B$_4$O$_{12}$ shows mechanical and thermodynamical stability with anisotropic nature
- Phonon properties of CP*B*O perovskite are investigated
- The optical properties related to the electric structure of CaPd$_3$B$_4$O$_{12}$ are investigated to justify the optoelectronic applications
- CPTO perovskite material has potential candidacy in thermoelectric applications.

**Graphical Abstract/TOC**

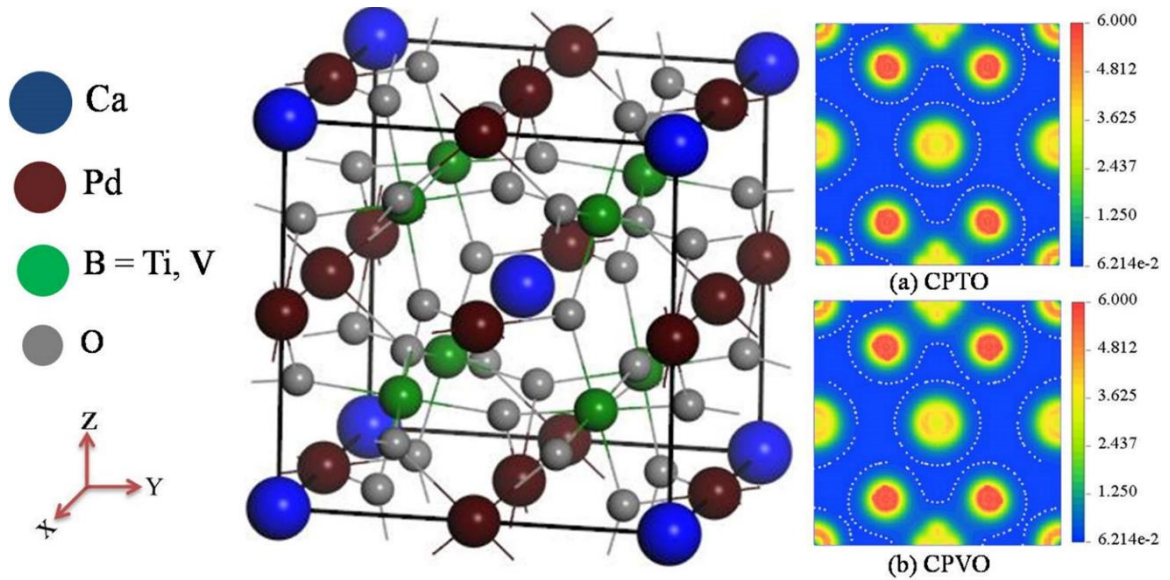

Crystal structure of cubic ($Im\bar{3}$) CaPd$_3$B$_4$O$_{12}$ (B = Ti, V) perovskites with the graphical representation of electronic charge density of (a) CaPd$_3$Ti$_4$O$_{12}$ and (b) CaPd$_3$V$_4$O$_{12}$.

**Abstract**

This study has explored numerous physical properties of $CaPd_3Ti_4O_{12}$ (CPTO) and $CaPd_3V_4O_{12}$ (CPVO) quadruple perovskites employing the density functional theory (DFT) method. The calculated lattice entities show inclinable compliance with the experimental results that ensure their structural stability. The mechanical permanence of these two compounds was observed by the Born stability criteria as well. The mechanical and elastic behaviors have been rationalized to investigate elastic constants, bulk, shear, and Young's modulus, Pugh's ratio, Poisson's ratio, and elastic anisotropy indexes. The ductility and anisotropic indexes confirm that both materials are ductile and elastically anisotropic in essence. The band structure of CPTO reveals a 0.88 and 0.46 eV direct narrow band gap while using GGA-mBJ and GGA-PBE potentials, respectively, which is an indication of its fascinating semiconducting nature. Whereas, CPVO perovskite exhibits a metallic character. The calculated partial density of states indicates the strong hybridization between Pd-4$d$ and O-2$p$ orbital electrons for CPTO, whereas Pd-4$d$ and V-3$d$-O-2$p$ for CPVO. The study of the chemical bonding nature and electronic charge distribution graph reveals the coexistence of covalent O-V/Pd bonds, ionic O-Ti/Ca bonds, as well as metallic Ti/V-Ti/V bonding for both compounds. The Fermi surface of CPVO ensures a kind of hole as well as electron faces simultaneously, indicating the multifarious band characteristic. The prediction of the static real dielectric function (optical property) of CPTO at zero energy implies its promising dielectric nature. The photoconductivity and absorption coefficient of CP*B*O display good qualitative compliance with the consequences of band structure computations. The calculated thermodynamic properties manifest the thermodynamical stability for CP*B*O, whereas phonon dispersions of CPVO exhibit stable phonon dispersion in contrast to slightly unstable phonon dispersion of CPTO. The predicted Debye temperature ($\theta_D$) has been utilized to correlate its topical features including thermoelectric behaviors. The studied thermoelectric transport properties of CPTO yielded the Seebeck coefficient (186 $\mu$V/K), power factor (11.9 $\mu$Wcm$^{-1}$K$^{-2}$), and figure of merit (ZT) value of about 0.8 at 800 K, indicating that this material could be a promising candidate for thermoelectric applications.

## 1 Introduction

Perovskite oxides (chemical formula $ABO_3$) and their various forms or modes are extensively studied owing to their diverse intriguing physical as well as chemical prominences. Studying $ABO_3$-type perovskite materials and their derivatives is essential for investigating phase transitions as well as understanding how different elements can be exchanged into their equivalent crystallographic positions to modify their structures and properties. Due to the exhibition of ferroelectric, multiferroic, magneto-electric, and superconducting characteristics, these materials are also fundamentally intriguing and have technological significance. These primarily induced properties are resulted because of interactivity within the transition metal cations in respect of B position and/or B–O–B interchange through oxygen anions. A large variety of chemical substitution is possible in perovskite compounds at both A and B cations sites [1–10]. In the midst of them, double and quadruple perovskites are considered as derivatives that have the general chemical formulas, $A_2BB'O_6$ or $AA'BB'O_6$ as well as $AA'_3BB'_4O_{12}$, respectively, where A and A′ indicate alkaline–earth and/or rare–earth metals whereas, B and B′ denotes transition metals. Now–a–days, such perovskites have been widely explored because of their diverse pompous physical as well as electrochemical attributes [1–6]. These types of perovskites are also tremendously used as magnetic memories, nonlinear optics, magneto-optic, magneto-ferroics, thermoelectric properties, tunnel junctions, and additional magnetic devices in the unprecedented spintronic arena [11,12]. An exceptional ordering in the A–site atoms results in an A-site ordered perovskite–structure oxide with the formula $AA'_3B_4O_{12}$ where the A and A′ ions are in the ordered state [13] is called quadruple perovskite also.

Among the A–site ordered perovskites two novel compounds $CaPd_3Ti_4O_{12}$ (CPTO) and $CaPd_3V_4O_{12}$ (CPVO) have been effectually fabricated by K. Shiro *et.* al., using high–pressure, high–temperature ambiance (15 GPa and 1000 °C) in which the A′–positions are fully occupied by rare $Pd^{2+}$ ions [14]. In this type of perovskites, three–quarters of the A–sites (A′–site) have a pseudo square planar coordination with Jahn–Teller active ions like $Cu^{2+}$ and $Mn^{2+}$, whereas the rest of A–positions are filled by ideal large A–site ions such as alkaline, alkaline earth, and rare earth metal ions [15]. In these structures transition–metal cations are set as twelve–fold coordinated A′–site creating square–coordinated units which are normal to each other. This aligning of the A′O$_4$ squares along with the large tilting of the corner–sharing $BO_6$ octahedra build the structural 2a×2a ×2a unit cell that deviates from a single perovskite structure. As these transition metal ions are positioned at the A′–site and the B–site, thus A′–A′ and A′–B interactions and further B–B interactions are observed in $ABO_3$ perovskite-type oxides. In recent years such interactions have drawn significant attention owing to their potential and useful properties [16,17]. However, currently, Rubel *et al.* and several researchers have explored Bi-based superconductive $AA′_3B_4O_{12}$–type perovskite structure [17–20] is termed as double perovskite as well on the basis of their unit cell volume. Furthermore, some theoretical studies have been implemented extensively on these Bi–oxide double and single perovskites [21–24].

The oxidation posture of synthesized $AA′_3B_4O_{12}$–type perovskite is confirmed by spectroscopic investigations. The authors also have reported the primal dependence of the metal-oxygen angles upon the deviation between the A′– and A–site ion sizes for the quadruple structural formula. Moreover, the measurements of the magnetic behavior, electrical resistivity, as well as specific heat ensured the diamagnetic insulating nature of CPTO, while CPVO was treated as Pauli–paramagnetic metal. Nonetheless, to our knowledge, no significant theoretical research works have been carried out to pursue additional important physical properties of CPTO and CPVO quadruple perovskites for device applications. DFT methods have been a promising and important technique in predicting numerous physical properties of solid materials for many years and more recently [25–29]. Interestingly, numerous articles [30–35] based on first-principles calculations on perovskite and other related structural compounds have been reported in order to accurately comprehend material systems and grasp the links between structure and property. Therefore, these two stable crystalline quadruple perovskites are our point of interest to investigate interesting and unexplored physical characteristics, by applying DFT schemes.

In the present study, the elastic, electronic, vibrational, as well as bonding behaviors of the above-mentioned compounds are computed on a most stable optimized structure using first-principle methods. However, in this investigation, crucial data on physical characteristics including band structure, elastic, optical, thermodynamic, or thermoelectric properties of CPVO and CPTO are calculated and analyzed to examine their potentiality for device applications. Among these physical peculiarities, thermoelectric properties are utilized for thermoelectric power generation which might be regarded as a long-term environment-friendly energy harvesting technology that can transmute the thermal power directly into electrical work and vice-versa with superior durability without producing noise or vibration. Since the thermoelectric conversion technology has drawn significant attraction, thus perovskite oxides with thermoelectric properties can potentially replace conventional thermoelectric materials because of their unique and comparatively simple form structure.

Nowadays, the thermoelectric performance of complex perovskite oxides for instance $Sr_{2-x}M_xB'MoO_6$: M = Ba, La, B' = Fe, Mn [36–38] has been studied from room to elevated temperature by several researchers those exhibited relatively better efficiency than that of conventional oxides. However, the conversion efficiency of a thermoelectric compound is evaluated in terms of entity figure of merit, $Z = (S^2\sigma/K)T$, where the parameters S, σ, K, and T denote the Seebeck coefficient, electrical conductivity, thermal conductivity, and absolute temperature, respectively. Notably, these perovskites are yet to employ practically with ZT < 1.0 condition. The band structure calculation of CPTO showed a direct narrow band gap that inspired us to perform thermoelectric properties of this material only. Moreover, the narrow band semiconductors which are made of p- and n-type carriers are considered promising candidates for thermoelectric device applications [21,39].

As far as we know, there is no report of temperature-dependent transport properties on the basis of theoretical and experimental methodology for CPTO quadruple perovskite. Therefore, the aforementioned scenarios have stimulated us to study various unexplored interesting behaviors of $CaPd_3Ti_4O_{12}$ and $CaPd_3V_4O_{12}$ perovskites based on DFT calculations by employing both CASTEP and WIEN2k programs. Consequently, we have calculated the structural, mechanical, electronic (band diagram, DOS, charge distribution with Fermi topology), optical, population analysis, thermodynamic, phonon, and thermoelectric properties of recently synthesized $CaPd_3B_4O_{12}$ (B = Ti, V) quadruple perovskites and calculated outcomes are compared with procurable results of the identical properties of materials.

## 2 Computational Method

The present first principle computations are executed by employing the well-known Cambridge Serial Total Energy Package (CASTEP) [25,40] and WIEN2k [41] programs in the framework of density functional theory (DFT) [42]. In this investigation, the fundamental set of the valence electronic states for Ca, Pd, Ti, V and O are

used as following sequence: $3s^23p^64s^2$, $4s^24p^64d^{10}$, $3p^63d^24s^2$, $3p^63d^34s^2$ and $2s^22p^4$, respectively. Initially, numerous functional or approximations such as GGA-PBEsol, GGA-PBE, GGA-RPBE, and GGA-TB-mBJ were applied to achieve the best-optimized structure. Among the various trial methods or approximations, the generalized gradient approximation (GGA) of Perdew–Burke–Ernzerhoffor (PBE) [43] and TB-mBJ [44] methods were finally selected, as these approximations matched well to gauge the electronic exchange with correlation potentials. The Vanderbilt–type ultrasoft pseudo potential [45] is also treated to represent the electrostatic interplay between valent electron and ionic core interactions. The energy cutoff ($E_{cut}$) of 700 eV is considered and a $k$–point mesh of 6×6×6 is employed for summation over the first Brillouin zone according to Monkhorst–Pack [46,47] scheme. Whereas, a relatively high-level $k$-point mesh 12×12×12 was utilized for the explicit visualization of the electronic charge density map and Fermi surface topology. The norm-conserving, as well as ultrasoft pseudopotential schemes of the GGA-PBEsol approach, is applied to calculate phonon dispersion and phonon density of states which deals with the interactions of valence electrons with ion cores. For the frequency-dependent phonon geometry optimization, the GGA-PBE approximation was employed and phonon dispersion properties were calculated using an $E_{cut}$ of 350 eV and 6×6×6 $k$- point mesh. We also apply the Broyden–Fletcher–Goldfarb–Shanno (BFGS) algorithm [48] for optimization of atomic order as well as density mixing is employed to optimize the electronic structure. Convergence tolerance for energy, maximum force, maximum displacement, and maximum stress is taken as $2\times10^{-5}$ eV/atom, 0.05 eV/Å, 0.02 Å, and 0.1 GPa, respectively, in the geometry optimization. The strain-stress method is very similar to that of previously reported metal carbide solids [49]. This procedure was used to calculate the elastic moduli (Cij) and mechanical characteristics of our studied cubic quadruple perovskites at ambient temperature and normal pressure (0.1 GPa) that did not pander the temperature cause. The electronic characteristics were also calculated by taking into account the same aforementioned parameters for both perovskites. We applied 0.5 Gaussian smearing to all computations in order to reduce the Fermi level and increase the effectiveness of the $k$-points on the Fermi surface. The calculations were thoroughly carried out without considering SOC for the diamagnetic and weak paramagnetic characteristics with complex structural scheme of materials. However, the electronic and thermodynamic properties of the materials are predicted via IRelast program interfaced with the WIEN2k code [50]. Various thermoelectric properties of CPTO oxide semiconductor have been computed through resolving Boltzmann semi-classical transport expressions as employed in BoltzTrap [51] interfaced with WIEN2k [41] program. In this method, a plane wave cut-off of kinetic energy $RK_{max} = 7.0$ is chosen to produce a good convergence for the self-consistent field (SCF) estimations. A mesh of 21×21×21 $k$-points was applied for calculating the thermoelectric behaviors. As GGA-PBE potential [52] debates the band gap of semiconductors, thus Tran-Blaha modified Becke-Johnson potential (TB-mBJ) [44] was also used in the calculation of electronic, transport, along with optical functions. In these calculations, the muffin tin radii 2.46, 2.09, 1.94, and 1.75 Bohr, respectively for Ca, Pd, Ti, and O were fixed in the calculations. During the computations of temperature-dependent transport properties, the value of chemical potential was fixed to that of Fermi energy. The parameter relaxation time (τ) is considered as a constant entity. The electronic conductivity as well as the electronic segment of thermal conductivity was accounted in terms of relaxation time, on the other hand the Seebeck coefficient was not dependent on this parameter.

## 3   Result and discussions

### 3.1   Structural properties

The two quaternary $CaPd_3Ti_4O_{12}$ and $CaPd_3V_4O_{12}$ perovskites crystallize in a cubic structure possessing the space group $Im\bar{3}$ (No. 204). The experimental equilibrium lattice parameters of $CaPd_3Ti_4O_{12}$ and $CaPd_3V_4O_{12}$ compounds are 7.49777 Å and 7.40317 Å, respectively [14]. In the optimized structures, the atomic position of Ca, Pd and Ti/V are (0, 0, 0), (0, 0.5, 0.5) and (0.25, 0.25, 0.25), respectively. Besides, the atomic positions of O were slightly distorted and are (0.2961, 0.1859, 0) and (0.2947, 0.1856, 0) respectively, for $CaPd_3Ti_4O_{12}$ and $CaPd_3V_4O_{12}$. The structure of these cubic compounds, $CaPd_3B_4O_{12}$ ($B$ = Ti, V) is depicted in **Figure 1**. **Table 1** represents the structure parameters and unit cell volume of titled $CaPd_3B_4O_{12}$ perovskites after the best compatible geometry optimization. It is seen that the achieved lattice parameters are well consistent with the available experimental values [14]. These resemble data of lattice parameters and might be originated from the replacement of Pd by Cu in $CaCu_3Ti_4O_{12}$ and $ACu_3Ru_4O_{12}$ (Ca, Sr) quadruples [14] and compared with reported double/quadruple perovskites [17,19,21,23].

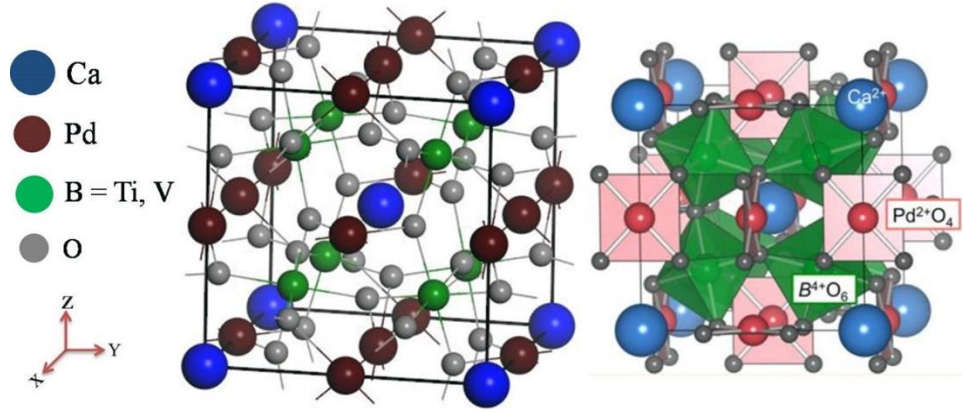

**Figure 1.** (Color online) Unit cell of cubic ($Im\bar{3}$) $CaPd_3B_4O_{12}$ ($B$ = Ti, V) perovskites (left−side) and three types of octahedra in the unit cell are shown by the blue (A), red (A′), green (B), and gray (O) spheres display the relevant atoms. The unit cell is built up of the atoms inside the solid lines: eight distorted $BO_6$ octahedra relative to one another, A′ ions bonded to four closest oxygen atoms, and A icons at the corners as well as at the body−center (right−side) (reproduced from Ref. [14]).

**Table 1.** Calculated lattice constants (*a*) in Å and unit cell volume (*V*) in Å$^3$ of $CaPd_3B_4O_{12}$ ($B$ = Ti, V) compounds.

| Compounds | Lattice constant, *a* (Experimental) [14] | Lattice constant, *a* (Present study) | Volume, *V* (Present study) |
|---|---|---|---|
| $CaPd_3Ti_4O_{12}$ | 7.50(14) [14] | 7.59 | 415.55 |
| $(NaK)Ba_3Bi_4O_{12}$ | 7.41(8) [14] | 7.39 | 395.27 |
| $(KBa)_3(Bi_{0.89}Na_{0.11})_4O_{12}$ | 8.50 [17] | 8.63 [21] | |
| | 8.53 [19] | 8.50 [23] | |

### 3.2 Mechanical and elastic properties

A structural solid's mechanical properties and elastic constants (Cij) show how it responds to external stress, and certain boundary factors are connected to a product's useful uses. The stiffness coefficients Cij of stable structural materials, which are elastic constants, provide us a thorough grasp of how the structural material reacts to applied stress between the elastic ranges. Additionally, a thorough understanding of the elastic constants of structural materials is essential for a variety of real-world applications related to the solid's mechanical properties, including thermoelastic stress, load deflection, internal strain, fracture toughness, and sound velocities. The nature of anisotropic indices, mechanical stability, and crystal symmetry directions are all highly correlated with the basic elastic and mechanical properties of solids. Only three self-sufficient independent elastic constants ($C_{11}$, $C_{12}$, and $C_{44}$) are required in order to fairly achieve the elastic and mechanical characteristics of cubic crystals like our examined materials. To show mechanical stability, the elastic constants of a cubic crystal must follow the Born criteria [42]: $C_{11} + 2C_{12} > 0$, $C_{11} > 0$, $C_{44} > 0$, $C_{11} - C_{12} > 0$. Three distinct elastic coefficients of both compounds are given in **Table 2**, which absolutely agree with the aforementioned functions, exhibiting the mechanical stability of $CaPd_3B_4O_{12}$ ($B$ = Ti, V).

**Table 2.** The elastic constants, $C_{ij}$ (GPa), bulk modulus, $B$ (GPa), shear modulus, $G$ (GPa), Young's modulus, $Y$ (GPa), Pugh's ratio, ($K = G/B$), Poisson's ratio, $v$, Vickers hardness, $H_v$ (GPa), Cauchy pressure (GPa), universal anisotropic index $A^U$, Burger's vector $b$ (Å), interlayer distance $d$ (Å) and Peierls stress $\sigma_P$ (GPa) of $CaPd_3B_4O_{12}$ compounds are compared with isostructural perovskites [21,23].

| Compounds | $C_{11}$ | $C_{12}$ | $C_{44}$ | $B$ | $G$ | $Y$ | $k = G/B$ | $v$ | $H_v$ | Cauchy pressure | $A^U$ | $b$ | $d$ | $\sigma_P$ |
|---|---|---|---|---|---|---|---|---|---|---|---|---|---|---|
| $CaPd_3Ti_4O_{12}$ | 310 | 210 | 141 | 243 | 93 | 248 | 0.38 | 0.33 | 6.13 7.58 | 69 | 0.15 | 7.59 | 3.79 | 1.51 |
| $CaPd_3V_4O_{12}$ | 404 | 160 | 82 | 242 | 96 | 255 | 0.40 | 0.32 | 6.88 8.22 | 78 | 0.19 | 7.39 | 3.69 | 1.39 |
| S1[24] | 223 | 48 | 53 | 107 | 65 | 162 | 0.61 | 0.25 | 9.8 | -5 | 0.55 | 8.63 | 4.31 | 2.6 |
| S2[26] | 241 | 64 | 49 | 123 | 63 | 161 | 0.51 | 0.28 | 7.3 | 15 | 0.55 | 8.50 | 4.26 | 1.1 |

S1 = $(NaK)Ba_3Bi_4O_{12}$, and S2 = $(KBa)_3(Bi_{0.89}Na_{0.11})_4O_{12}$

The remaining other polycrystalline elastic constants/coefficients such as the bulk modulus ($B$), Young's modulus ($Y$), shear modulus ($G$), and Poisson's ratio ($v$) are determined using $C_{ij}$ via the Voigt ($V$)−Reuss ($R$)−Hill ($H$) scheme [53]. All the elastic parameters are listed in **Table 2** and compared with isostructural perovskites [21,23]. The $B$ is a function that is useful for evaluating a material's average bond potency inside the atoms [42]. The calculated values of $B$ (243 GPa and 242 GPa for $CaPd_3Ti_4O_{12}$ and $CaPd_3V_4O_{12}$, respectively) confirm the strong bonding of atoms within these compounds. The bonding strength also offers the necessary resistance to volume deformity caused by the action of externally applied pressure. Besides, any deflection in the form of a solid largely relies upon $G$, which shows a crucial correlation with the material's hardness. To estimate the material's hardness directly, the Vickers hardness parameter is frequently used and defined by: $Hv = 2(k^2 G)^{0.585} - 3$ [54] and the also equation employed by Liu et. al. [49], $H_V = 0.92 K^{1.137} G^{0.708}$, where $k$ ($= G/B$) is the Pugh's ratio. The calculated values of the Pugh's ratio ($k$), and Vickers hardness ($Hv$) using both equations, Poisson's ratio ($v$), and Cauchy pressure are given in **Table 2**. Diamond has been determined to be the hardest substance with a Vickers hardness ranging from 70 to 150 GPa. The calculated $Hv$ for CPTO are 6.13 and 7.58 GPa, whereas, those for CPVO are 6.88 and 8.22 GPa (**Table 2**), respectively. Although, the same parameters are used for the $H_V$ calculations the meager difference between these two magnitudes of hardness might be responsible for the variation of critical exponent and constants in the expressions. Therefore, we can conclude that the perovskites being studied are comparatively very mild in contrast to diamonds but almost comparable to reported superconducting perovskites [20–24].

In most practical applications, a compound is needed to identify it as either ductile or brittle. To become ductile a material must have a $k$ value lower than the critical value of 0.57, while a higher $k$ value than the critical one reveals the material's brittle nature [55]; The $k$ values of $CaPd_3Ti_4O_{12}$ and $CaPd_3V_4O_{12}$ are found to be 0.40 and 0.38 (**Table 2**), respectively, indicating the ductile nature of them. The parameter Poisson's ratio ($v$) indicates the bonding and ductile characteristics of the solids. If its value is around 0.1, implies a covalent compound, whereas to be an ionic material $v$ must be greater than 0.25 [56]. From the calculation, it is very clear that both quadruple have ionic bonding but CPVO is slightly ionic than CPTO. Further, Frantsevich et. al. also established a critical value of Poisson's ratio ($v = 0.26$) in order to comprehend the ductile character of solids [57]. The values of calculated $v$ for $CaPd_3Ti_4O_{12}$ and $CaPd_3V_4O_{12}$ are 0.31 and 0.32 (**Table 2**), which also predict the ductility of both perovskites. Moreover, Cauchy pressure is another indicator to identify the ductile/brittle behavior of a solid which is expressed as ($C_{12}$−$C_{44}$) [58]. If the Cauchy pressure of material has a negative value, it is supposed a brittle characteristic[24]; otherwise (having a positive value) demonstrates the ductile behavior of that material [58]. Since the Cauchy pressure is positive for both compounds as seen in **Table 2**, they possess ductile nature and metal bonding as well. Therefore, the compounds $CaPd_3B_4O_{12}$ in this study show ductile characteristics considering all three indicatives.

The universal anisotropic index, denoted by $A^U = 5\frac{G_V}{G_R} + \frac{B_V}{B_R} - 6 \geq 0$, is an indicator of crystal anisotropy that has been used to calculate the anisotropic nature of the materials. Material is fully isotropic if its $A^U = 0$, but the deviation from zero discloses the level of elastic anisotropy [59]. The magnitudes of $A^U$ for $CaPd_3Ti_4O_{12}$ and $CaPd_3V_4O_{12}$ are 0.15 and 0.19 (**Table 2**), respectively implying significant anisotropy of these materials. On the other hand, the isostructural perovskites [21,23] presented in **Table 2** show a very high level of anisotropy.

In this study, we also have calculated Peierls stress ($\sigma_p$) to predict the strength of $CaPd_3B_4O_{12}$ for motile a dislocation inside the atomic plane by using the expression [42] as follows (**Eq. 1**):

$$\sigma_p = \frac{G}{1-v} \exp\left[-\frac{2\pi d}{b(1-v)}\right] \quad (1)$$

Herein, $G$, $v$, $b$ and $d$ are the shear modulus, Poisson ratio, Burgers vector, and interplanar length within the glide planes (**Table 2**). The calculated $\sigma_p$ of $CaPd_3B_4O_{12}$ is found to be 1.51 and 1.39, respectively for CPTO and CPVO. These values of $\sigma_p$ of $CaPd_3B_4O_{12}$ in this study are compared with some perovskite superconductors[23-26], inverse perovskites $Sc_3InX$ (X = B, C, N), MAX phases, and rock salt binary carbides [60] those are exhibiting the sequence: $\sigma_p$ (chosen perovskites, inverse perovskites as well as MAX phases) $<\sigma_p$ ($CaPd_3B_4O_{12}$ quadruple perovskites) $<<\sigma_p$ (binary carbides). Therefore, it is demonstrated that dislocations can move in the approved MAX phases easily, whereas it is quite impossible for binary carbides. As $\sigma_p$ of $CaPd_3B_4O_{12}$ possesses between MAX phases and binary carbides, dislocation can still move, but not so easily as in MAX phase compounds. However, dislocation movement in CaPd$_3$V$_4$O$_{12}$ may occur more easily for a lower value of $\sigma_p$ in comparison to CaPd$_3$Ti$_4$O$_{12}$.

**3.3. Electronic properties**

*3.2.1 Band structure (BS) and density of states (DOS)*

The electronic band diagram, the partial density of state (PDOS), as well as total density of state (TDOS) of CaPd$_3$B$_4$O$_{12}$ are depicted in **Figures 2-4**, respectively, employing both GGA-PBE and TB-mBJ potentials for comparison between the methods. It is well known and reported that GGA-PBE debates bandgaps [61,62] in contrast to TB-mBJ and other hybrid approaches like HSE06 or DFT+U+SOC etc. The electronic band structure confirms either material is a conductor, semiconductor, or insulator. The PDOS and TDOS are defined in terms of a number of states at a filled or unoccupied energy equality in statistical as well as condensed matter physics [63]. It indicates hybridization in the midst of orbital electrons and bonding nature amid the compounds. The Fermi level, $E_F$ is represented using a horizontal broken streak. In the band diagram, the pure valence and conduction bands are shown by magenta and blue lines respectively, whereas the bands crossing the $E_F$ are indicated with orange lines. As can be seen from the band diagram of **Figure 2(a)**, there is no overlap between the valence and conduction bands and exhibits a direct band gap, $E_g$~0.46 eV for the CPTO while employing a GGA-PBE potential. The PDOS of CPTO reveals robust hybridization between Pd–4$d$ and O–2$p$ orbitals at −3.0 eV to $E_F$ for both materials as presented in **Figure 2(b)**. The bands of DOS originate from the contribution of O–2$p$ orbital electrons in this energy range which is higher than Ca–4$p$/4$s$ and Ti–3$p$/3$d$ orbital electrons. Notably, in the midst of all orbitals Pd–4$d$ state near the $E_F$ contributes predominately to the PDOS and hence TDOS for CaPd$_3$Ti$_4$O$_{12}$. It is well-known that the GGA-PBE functional usually lowers the band gap of semiconductors owing to its average potential consideration. Thus, the more localized TB-mBJ potential is also utilized to further accurately predict the band gap with valence and conduction band dispersions of CPTO [61,62] in the band diagram. A direct bandgap, $E_g$ ~0.88 eV, and smeared band structure are determined using this potential as displayed in **Figure 3(a)**. This result suggests that the attained band gap of CPTO using TB-mBJ shows good consistency with the empiric outcomes [14]. Importantly, the valence bands adjacent to $E_F$ are smoother in contrast to conduction bands for this material using both potentials. The flatness of these bands originates from the robust hybridization of Pd–4$d$ and O–2$p$ states which are also reflected in the PDOS [39] of **Figure 3(b)**. The total DOS at $E_F$ for this compound is approximately 3.77 states/eV/unit cells is an indication of its conductive nature of it as well. Moreover, the obtained electronic band structure and density of states (DOS) of CaPd$_3$V$_4$O$_{12}$ are displayed in **Figure 4(a-b)** using GGA approximation. It is noticed from **Figure 4(a)** that some of the valence bands crossed the $E_F$ and overlapped with conduction bands, which is strong evidence of metallic behavior [14] of the CPVO quadruple perovskite. This finding is consistent with some reported studies [21,22,64,65] as well. On the other hand, from **Figure 4(b),** we see the strong hybridization among Pd–4$d$, V–3$d$, and O–2$p$ orbital electrons maintaining an energy scale of −3.0 eV to $E_F$. At the $E_F$ the most dominant contribution comes from V–3$d$ orbital electrons in comparison to Pd–4$d$ and O–2$p$ states. Herein, the contribution of Ca–4$s$/4$p$ states is remarkable at $E_F$ but exhibits a small value than other orbitals. However, the individual share of O–2$p$ orbital in the vicinity of $E_F$ is very similar to the common characteristics of PDOS of reported double and simple perovskite compounds [13,66]. Noteworthy, the strong hybridization among Pd–4$d$, V–3$d$ and O–2$p$ states imply a strong ternary ionic/covalent Pd-V-O bond in CaPd$_3$V$_4$O$_{12}$ perovskite. Therefore, the electronic structure calculations of CPVO demonstrate that the A′-site ions keep an important role in the metallic conduction of V−based perovskites [14].

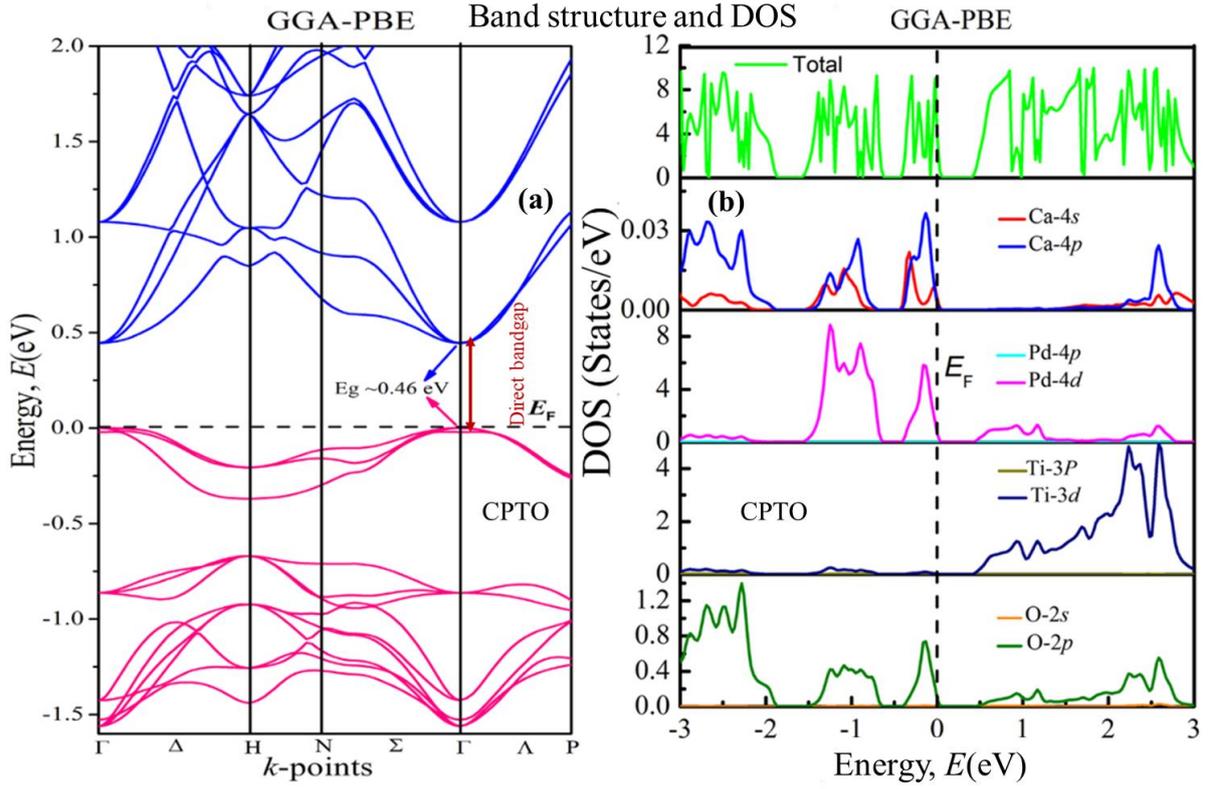

**Figure 2.** (a) Calculated band structure, (b) partial density of states, and TDOS of $CaPd_3Ti_4O_{12}$ using GGA–PBE potentials along with the exalted symmetric approaches of the Brillouin zone. In the figure, the $E_F$ is referred to as the valence band maxima (VBM).

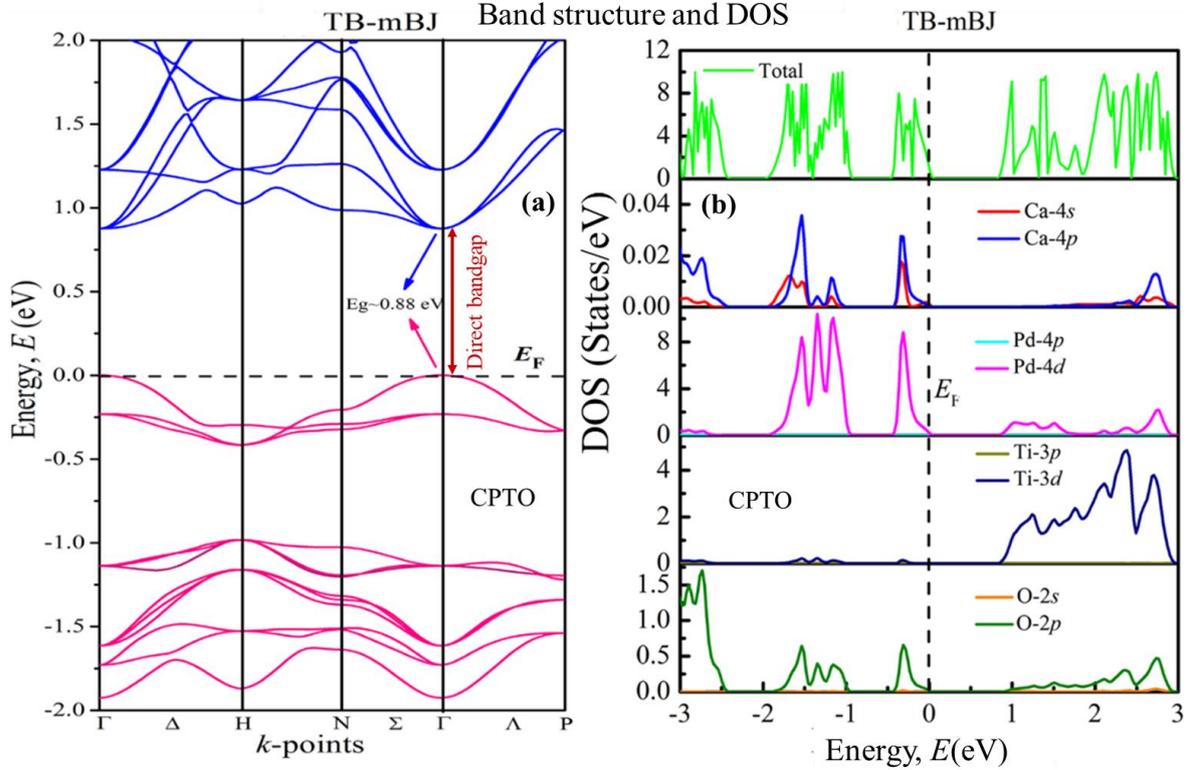

**Figure 3.** (a) Calculated band structure, (b) partial density of states, and TDOS of $CaPd_3Ti_4O_{12}$ using TB-mBJ potential along with the exalted symmetric approaches of the Brillouin zone. In the figure, the $E_F$ is referred to as the valence band maxima (VBM).

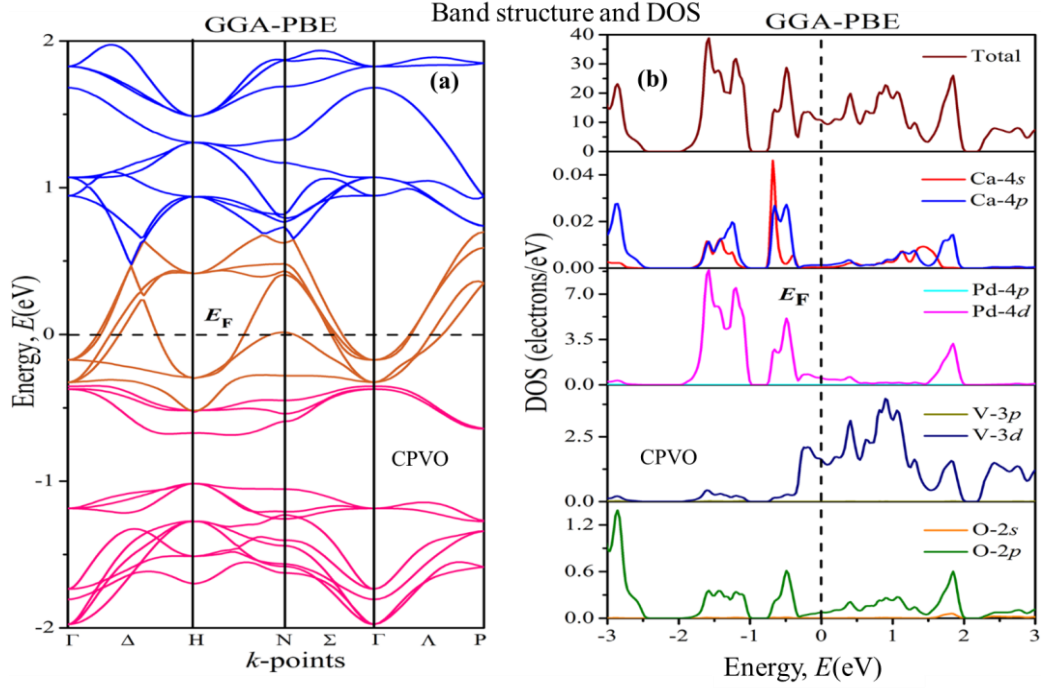

**Figure 4.** (a) Calculated band structure, (b) partial density of states as well as TDOS of $CaPd_3V_4O_{12}$ using GGA-PBE approximation along the same symmetry points of CPTO in the Brillouin zone.

*3.2.2    Electronic charge distribution and fermi surface*

The study of electronic charge density is a supporting portion to analyze the electronic property more significantly of a material system. This feature yields a charge density structural map of valence electrons regarding the total distribution of charge in a unit cell of a compound. The behavior of chemical bonding within a compound is also analyzed by investigating the total electronic charge density map. Basically, the charge density map/curve consists of structural atoms/ions indicating the contribution of orbital electrons in the electronic properties through accumulating charges of atoms/ions. Subsequently, the distinguished color density map correlates the electronic DOS spectra of constituent elements by introducing their charge contributions. **Figures 5(a)−(b)** present the valence electronic charge density (in the units of $e/Å^3$) map of $CaPd_3B_4O_{12}$ double perovskite along the (100) plane. The adjacent scale on the right side of both contour plots of **Figures 5(a)-(c)** reveals the acuity of charge (electron) density. The high and low intensity are imparted by blue and red colors, respectively. The Pd−O bond in $CaPd_3Ti_4O_{12}$ corresponds to the hybridization of Pd−4$d$ with O−2$p$ orbitals as seen in **Figure 3**. Moreover, we depicted the band decomposed electronic charge density maps by taking only two and four bands adjacent to conduction band maxima (CBM) and valence band maxima (VBM) of the $CaPd_3Ti_4O_{12}$ semiconductor (**Figure 5(b)**). From **Figure 5(b)** it is clear that the O-2$p$ orbital has dominant charge density distribution in contrast to other orbitals in the CBM and VBM at $E_F$ of electronic band structure (**Figure 2**). Whereas Pd−V−O in $CaPd_3V_4O_{12}$ corresponds to the vigorous hybridization of Pd−4$d$, O−2$p$ and V−4$d$ orbitals is evident from DOS in **Figure 4**. From crystal structure, it is also assumed that (Ti/V = $B$)O$_6$ octahedra is built through ionic bonding. We see that O and Pd ions reveal ionic characteristics in the plane (100) even though the charge density contour of oxygen is not entirely spherical but it shows ionic bonding. Moreover, the existence of covalency between Ca and Pd atoms is observed along with Pd and O ionic bonds. The variation in electronegativity of subsisting ions in the crystal lattice is favorable for such types of bondings. For the CPTO compound, the Ti-O bond is covalent and the Ti-Ti bond shows a metallic character while for CPVO solid V-V bond is metallic with a V-O covalent bond, and this is higher covalency of V and O might be liable for the metal character of CPVO. Further, the charge density arrangement surrounding all atoms (excluding oxygen) is likely spherical which shows ionic nature and is evident from Pd−O and Pd−V−O bonds of CPVO. This ionic nature is caused by metallic characteristics [67] as well though the band diagram of CPTO manifests the characteristics of a degenerate semiconductor. Therefore, a combination of chemical bonding namely ionic, covalent and metallic interactions exists within $CaPd_3B_4O_{12}$. Since the O atom has higher electronegativity in contrast to others, the charges gather dominantly near the O atom.

The calculated electronic charge density mapping along various planes produced a similar outcome is an indication of their isotropic nature as well. Furthermore, the calculated bond valence sum (BVS) of CPTO indicated the ionic model as $Ca^{2+}Pd^{2+}_3Ti^{4+}_4O_{12}$ whereas that for CPVO implies $Ca^{2+}Pd^{2+}_3V^{4+}_4O_{12}$. Since the charge density contour is roughly spherical around each atom for both compounds, it reflects the ionic nature of Pd–O and Pd–V–O bonds. It is seen that ionic as well as covalent bonds are robust in CPVO compared to those of CPTO.

For being a diamagnetic insulator/semiconductor, we could not calculate the Fermi surface topology of $CaPd_3Ti_4O_{12}$. On the other hand, the Fermi surface of $CaPd_3V_4O_{12}$ is calculated by taking the $E_F$ as valence band maxima (VBM) for the bands crossing $E_F$ as shown in **Figure 5(c)**. It is noticed from the figure that an electron–like a sheet with a cubic cross-section is centered toward the Γ−Z path of the Brillouin zone. There is also a hole–like a Fermi surface around X–point. At the nook of the Brillouin zone, another electron–like the Fermi surface around R–point is connected with the hole-like Fermi surface. Therefore, it is evident that both electron and hole–like Fermi surfaces exist in CPVO which indicates the multiple–band nature of this compound.

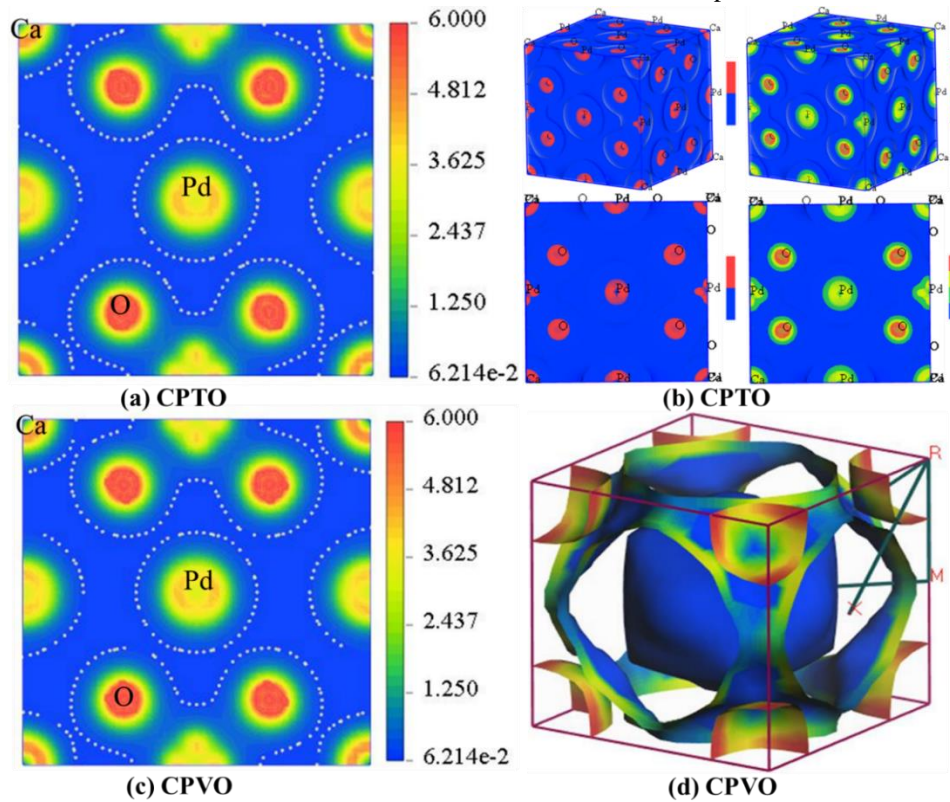

**Figure 5.** Electronic charge density of (a-b) $CaPd_3Ti_4O_{12}$ and (c) $CaPd_3V_4O_{12}$. The Fermi surface topology of $CaPd_3V_4O_{12}$ double perovskite is depicted in (d).

### 3.3 Optical properties of CPTO and CPVO

The optical properties of perovskite materials play a significant role in the field of optoelectronics as nanoelectronics devices [68] are very effective to explain the electronic structure of solids. The various optical functions have been calculated to inspect the electronic structure and feasible applications of $CaPd_3B_4O_{12}$. In the midst of numerous optical functions, the real and imaginary parts of the dielectric function are obtained as a function of applied optical frequencies (energy). The real part of the dielectric function indicates or measures the amount of dielectric polarization or constants of a material under an external frequency. On the other hand, the imaginary part of the dielectric function/constant estimates the dielectric losses by absorption of energy (electromagnetic radiations/photons) as atomic or molecular vibrations. The lower the imaginary dielectric function and higher the real dielectric function. If the imaginary dielectric function is zero or negative at the low-frequency range it means no dielectric losses at that incident frequency (**Figure 6**).

The real ($\varepsilon_1$) as well as imaginary ($\varepsilon_2$) sections of the dielectric function precisely describe the optical behaviors of materials [69] and have been determined using the frequency-dependent dielectric expression, $\varepsilon(\omega)$

= $\varepsilon_1(\omega) + i\varepsilon_2(\omega)$ which has a direct relation to the electronic configurations of materials. **Figures 6(a)** and **6(b)** depict the spectra of $\varepsilon_1$ and $\varepsilon_2$ with respect to photon energy for CaPd$_3$B$_4$O$_{12}$. The real part of the dielectric function at zero frequency limits is called the electronic part of the static dielectric function and it is found to be 11.07 for CaPd$_3$Ti$_4$O$_{12}$ (**Figure 6(a)**). However, $\varepsilon_1$ goes below zero at around 8.17 eV and back to zero at about 12.18 eV. Notably, the negative value of the $\varepsilon_1$ implies the Drude–like the behavior of CaPd$_3$Ti$_4$O$_{12}$. The $\varepsilon_2$ approaches zero around 40.64 eV. Besides, CaPd$_3$V$_4$O$_{12}$ is a metallic material that is clearly seen from the band diagram (**Figure 4**). Owing to the metallic nature of CPVO, we employ both the Drude plasma frequency along with a damping factor of 3.0 eV and 0.05 eV, respectively, to determine its dielectric function. The peak of $\varepsilon_1$ is connected to the electron excitation and is mainly arisen due to the intraband transitions generated mainly from low-energy conduction electrons. Significantly, the $\varepsilon_2(\omega)$ approaches zero at around 30 eV in the ultraviolet frequency part, suggesting the transparent and anisotropic nature of CPVO. The analysis of elastic parameters also reflects the anisotropic nature of CPVO.

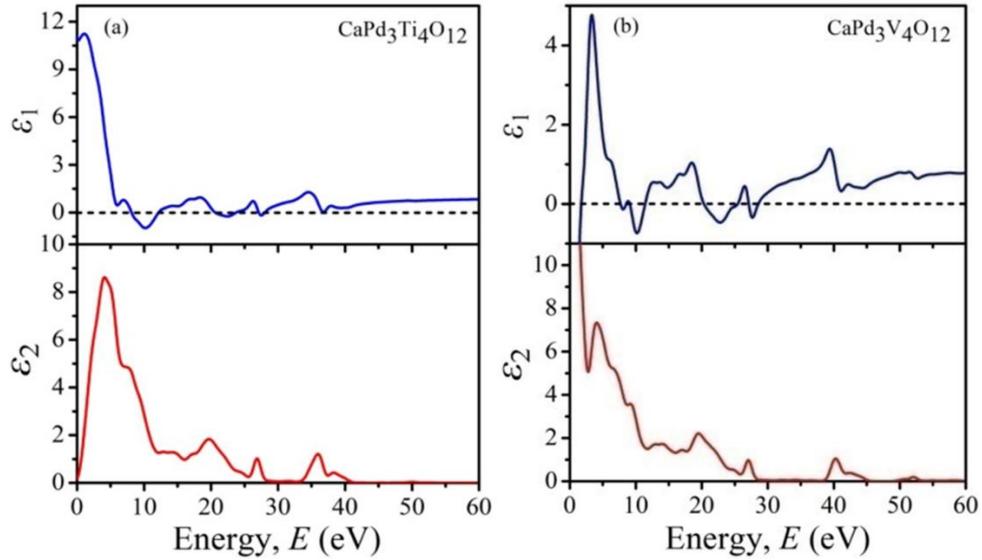

**Figure 6.** Frequency dependence dielectric function with $\varepsilon_1$, and $\varepsilon_2$ of (a) CaPd$_3$Ti$_4$O$_{12}$ and (b) CaPd$_3$V$_4$O$_{12}$.

A material to be used in photovoltaic systems with high optical conductivity, high absorption, high value of the refractive index, and less emissivity is crucial and required condition. Among these, the optical compatibility of a material is usually estimated based on the sense of refractive index to be applied in optical devices together with waveguides, data storage media, photonic crystals, and so on [70]. When an electromagnetic wave passes through a substance, the refractive index indicates the quantity of absorption loss. The refractive index ($n$), as well as the extinction coefficient ($k$) of CaPd$_3$B$_4$O$_{12}$, are shown in **Figure 7**.

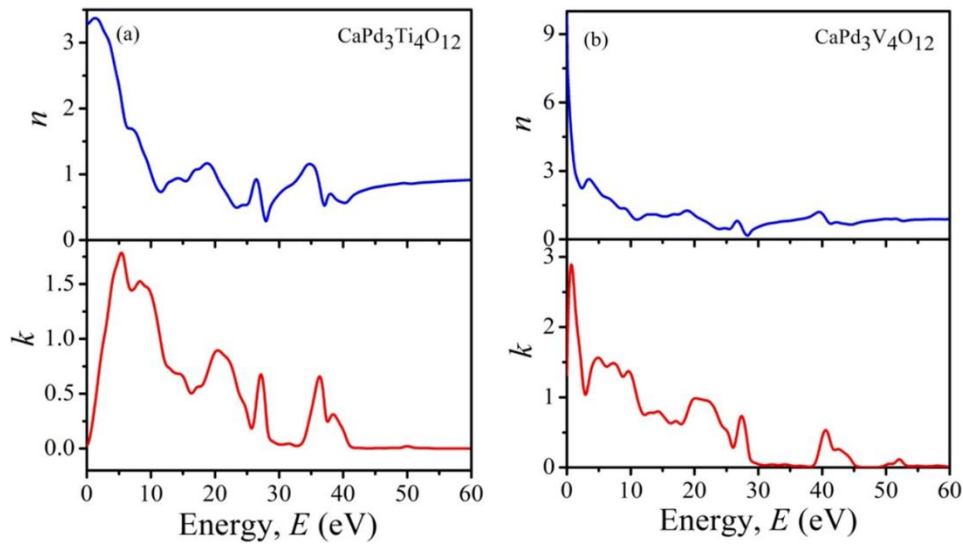

**Figure 7.** (Color online) The refractive index ($n$) and the extinction coefficient ($k$) nature of (a) $CaPd_3Ti_4O_{12}$ and (b) $CaPd_3V_4O_{12}$.

The calculated values of the static refractive index $n(0)$ of CPTO and CPVO are 3.32 (**Figure 7(a)**) and 9.78 (**Figure 7(b)**), respectively, and decreases in the upper-frequency portion as well. The predicted $n(0)$ of $CaPd_3Ti_4O_{12}$ is a little higher than that of the $K_2Cu_2GeS_4$ semiconductor but in the vicinity of GaAs semiconductor [39].

**Figure 8(a)** shows the real part of conductivity spectra ($\sigma$) of $CaPd_3B_4O_{12}$. It is evident that the photoconductivity of $CaPd_3Ti_4O_{12}$ has not initiated at zero frequency due to its small band gap which can also be seen from the electronic band diagram. On the other hand, the photoconductivity of the $CaPd_3V_4O_{12}$ compound starts from zero photon energy owing to its metallic characteristics. However, the maximum photo conductivities are 5.25 and 5.23 found at 35.96 and 19.58 eV for $CaPd_3Ti_4O_{12}$ and $CaPd_3V_4O_{12}$, respectively.

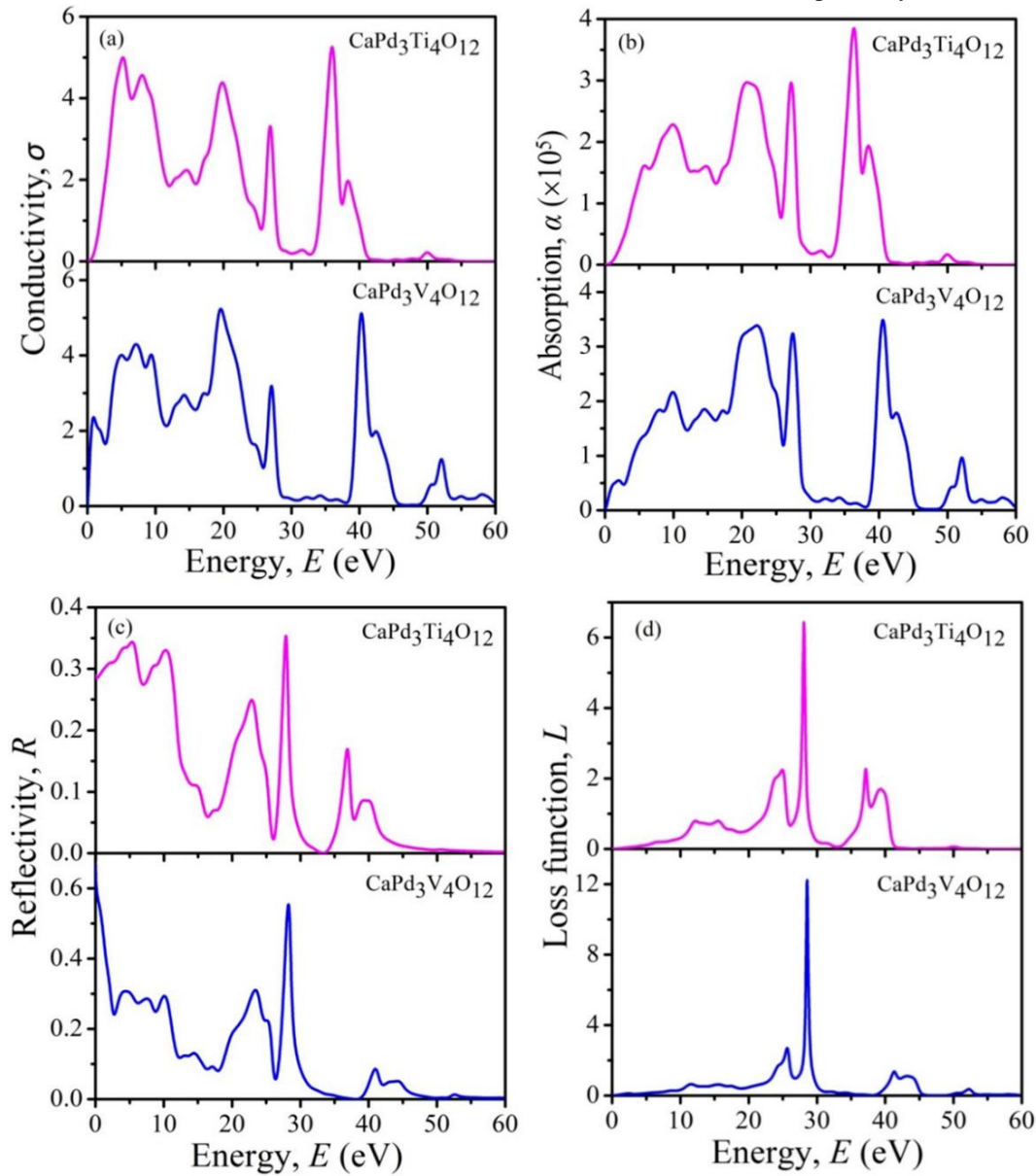

**Figure 8.** Energy-dependent of **(a)** real part of conductivity ($\sigma$), **(b)** absorption ($\alpha$), **(c)** reflectivity ($R$), and **(d)** loss function ($L$) of $CaPd_3B_4O_{12}$.

It is noteworthy, that the photoconductivity and thus electrical conductivity of a semiconductor enhances with absorbed photons [71]. **Figure 8(b)** represents the absorption coefficient spectra ($\alpha$) of $CaPd_3B_4O_{12}$ that are related to the optimal solar energy altering efficiency which implies the percolation capability of materials before being absorbed. It is visualized that the absorption for $CaPd_3Ti_4O_{12}$ begins ascending at ~0.90 eV owing to the semiconducting behavior together with the calculated band gap of 0.88 eV which is acquainted with absorption edges as well. Whereas, the absorption of $CaPd_3V_4O_{12}$ metal starts increasing from zero energy because of overlapping of bands (metallicity) as seen from band structure calculation. However, generally, the intraband contribution mainly affects the low-energy infrared part of the spectra. In addition, the peaks in the high-frequency region $α$ and $σ$ may originate because of the interband transition. Therefore, it is very clear from **Figures 8(a)** and **(b)** that the change of $σ$ is almost identical to that of $α$ for both these compounds. Hence, the photoconductivity of $CaPd_3B_4O_{12}$ rises due to photon absorption [72]. The reflectivity spectrum is a crucial optical parameter to determine all the optical constants using Kramers-Kroning relations and the reflectivity spectra ($R$) for $CaPd_3B_4O_{12}$ are depicted in **Figure 8(c)**. It is observed that the $R$ spectra in the case of both compounds initiated at zero energy are regarded as the static portion of reflectivity. Interestingly, a moderate reflectivity of approximately 28% for $CaPd_3Ti_4O_{12}$ but a very high reflectivity of ~66% is seen for $CaPd_3V_4O_{12}$ in the infrared region. Moreover, some peaks are found for both materials in the superior-energy region for interband transitions. The energy loss spectrums ($L$) of $CaPd_3B_4O_{12}$ are illustrated in **Figure 8(d)**. It is an important parameter that reveals how much energy a fast electron loses during passing through a substance [73]. The highest loss peak is related to the plasma resonance and its accompanied frequency is said to be plasma frequency $ω_p$ [74]. The maximum peaks are found at 28.07 and 28.57 eV for $CaPd_3Ti_4O_{12}$ and $CaPd_3V_4O_{12}$ respectively, revealing the plasma frequency of these two perovskites.

### 3.4 Population analysis

Relatively vast interesting information is obtained from Mulliken atomic population analysis which is needed for understanding material's bonding behavior [75]. The obtained data from population analysis are summarized in **Table 3** and **Table 4**. If the bond overlap population is zero, the bond is expected to be ionic, whereas a value higher than zero indicates the incremental degree of covalent nature [76].

**Table 3.** Mulliken atomic population inspection of $CaPd_3B_4O_{12}$ compounds.

| Compounds | Species | Mulliken atomic populations | | | | Charge (e) |
|---|---|---|---|---|---|---|
| | | s | p | d | Total | |
| $CaPd_3Ti_4O_{12}$ | Ca | 2.10 | 5.99 | 0.59 | 8.67 | 1.33 |
| | Pd | 2.53 | 6.17 | 8.77 | 17.47 | 0.53 |
| | Ti | 2.28 | 6.49 | 2.14 | 10.91 | 1.09 |
| | O | 1.83 | 4.77 | 0.00 | 6.60 | -0.60 |
| $CaPd_3V_4O_{12}$ | Ca | 2.10 | 5.99 | 0.60 | 8.70 | 1.30 |
| | Pd | 2.50 | 6.13 | 8.82 | 17.45 | 0.55 |
| | V | 2.30 | 4.48 | 3.24 | 12.03 | 0.97 |
| | O | 1.83 | 4.74 | 0.00 | 6.57 | -0.57 |

**Table 4.** Computed Mulliken bond number $n^μ$, bond length $d^μ$ and bond overlap population $P^μ$, of $CaPd_3B_4O_{12}$ solids.

| Compounds | Bonds | $n^μ$ | $d^μ(Å)$ | $P^μ$ |
|---|---|---|---|---|
| $CaPd_3Ti_4O_{12}$ | O–Pd (I) | 24 | 2.02 | 0.27 |
| | O-Pd (II) | 24 | 2.85 | -0.20 |
| | O-Ti (III) | 48 | 1.97 | 0.48 |
| $CaPd_3V_4O_{12}$ | O–Pd (I) | 24 | 2.04 | 0.25 |
| | O-Pd (II) | 24 | 2.77 | -0.19 |

| | | | |
|---|---|---|---|
| O–V (III) | 48 | 1.92 | 0.49 |

From **Table 4** it is seen that the atomic bond populations for both compounds are positive and negative. A negative value indicates ionic nature but a value higher than zero reveals the covalency of these compounds. Therefore, we can conclude that both ionic and covalent bonds exist within $CaPd_3B_4O_{12}$ which agrees with the electronic charge density map. However, the covalent nature of the O-Ti bond in $CaPd_3Ti_4O_{12}$ is slightly weaker compared to the O−V bond in $CaPd_3V_4O_{12}$ which is also stronger than that of other existing bonds in $CaPd_3B_4O_{12}$.

### 3.5  Thermodynamic properties and phonon dispersions

Various thermodynamic attributes for instance Debye temperature ($\theta_D$), melting temperature ($T_m$) and thermal conductivity of the $CaPd_3B_4O_{12}$ perovskites are calculated to understand their behavior under high temperatures and high pressures. Debye temperature ($\theta_D$) plays an important role to address some interesting physical properties namely lattice vibrations, thermal conductivity, minimum thermal conductivity, melting point, specific heat, and so on. Interestingly, $\theta_D$ is connected to the lattice thermal conductivity to evaluate the thermoelectric performance of a material. Thus, it is noteworthy to calculate the Debye temperature of these perovskites to have an idea respecting thermal conductivity via the expression as follows (**Eq. 2** [77]);

$$\theta_D = \frac{h}{k_B}\left[\frac{3m}{4\pi}\left(\frac{N_A \rho}{M}\right)\right]^{\frac{1}{3}} v_m \qquad (2)$$

where, $h$ and $k_B$ are the Planck's and Boltzmann constants, respectively. $V$ denotes the volume of the unit cell, $n$ indicates the number of atoms within a unit cell, and $v_m$ implies the average sound velocity. The $v_m$ can be determined using the expression below (**Eq. 3**),

$$v_m = \left[\frac{1}{3}\left(\frac{2}{v_t^3} + \frac{1}{v_l^3}\right)\right]^{-\frac{1}{3}} \qquad (3)$$

Here, $v_l$ and $v_t$ represent the longitudinal and transverse sound velocities, respectively. By using the value of bulk modulus, $B$ and shear modulus, $G$, the $v_l$ and $v_t$ can be estimated using the following expression (**Eq. 4**),

$$v_l = \left(\frac{B+\frac{4}{3}G}{\rho}\right) \text{ and } v_t = [G/\rho]^{1/2}. \qquad (4)$$

The melting temperature of a crystal is an essential entity for the application in a heating system that can be calculated employing the expression [78] as follows (**Eq. 5**),

$$T_m = \left[553\,K + \left(\frac{5.91\,K}{GPa}\right)C_{11}\right] \pm 300\,K \qquad (5)$$

In cubic structure, the axial lengths are equal; therefore, the elastic constants $C_{11}$, $C_{22}$, and $C_{33}$ are also equal. The calculated melting temperature ($T_m$) of CPVO and CPTO materials is high (**Table 5**) which makes them favorable for elevated temperature applications. All the calculated values of $\theta_D$, $T_m$, $v_m$, $v_t$, and $v_l$ $K_{min}$ for the titled compounds are listed in **Table 5**. The calculated melting temperature ($T_m$) of CPVO and CPTO materials is high (**Table 5**) which makes them favorable for elevated temperature applications. From **Tables 2** and **5** it is seen that the relatively high value of $\Theta_D$, $T_m$, and elastic constants (B, G, and Y) imply the dynamical stability as well as the hard nature of the perovskites [66]. The term minimum thermal conductivity ($K_{min}$) is straight way related to the temperature factor and Debye temperature ($\Theta_D$) which consequently generates lattice vibration in the materials. The increment of the temperature inchmeal rises the lattice conductivity to a particular limit and then again extenuates to a fixed limit [66]. The minimum thermal conductivity, $K_{min}$ has been computed by making use of familiar Clarke equalization in the following [79]:

$$K_{min} = k_B v_m (M/n\rho N_A)^{-2/3} \qquad (6)$$

Herein, $k_B$ denotes the Boltzmann constant, $v_m$ denotes the average sound velocity, M is the molecular mass, n indicates the number of atoms per molecule, and $N_A$ expresses Avogadro's number. From the above equation, it is clear that $K_{min}$ depends only on sound velocity or phonon/vibrational frequency. Thus, the $K_{min}$ attributes the lattice thermal conductivity owing to vibrational modes (phonons). Moreover, the entity $\Theta_D$ also depends on the temperature and/or pressure effects of phonon or vibrational mode of materials which can be calculated using PHDOS. Therefore, a strong correlation between $K_{min}$ and $\Theta_D$ is observed for the lattice thermal conductivity in the materials. However, CPVO exhibits slightly superior electronic thermal conductivity in contrast to CPTO because of its metal character. This behavior of thermal conductivity can be utilized to search utmost as well as minimum thermal appliances near future.

**Table 5.** The estimated density ($\rho$), longitudinal, transverse, and average sound velocities ($v_l$, $v_t$, and $v_m$) Debye temperature ($\Theta_D$), melting temperature ($T_m$), and minimum thermal conductivity ($K_{min}$) of $CaPd_3B_4O_{12}$ quadruple perovskite.

| Compounds | $\rho$ (g/cm$^3$) | $v_l$(km/s) | $v_t$(km/s) | $v_m$(km/s) | $\Theta_D$(K) | $T_m$ (K) | $K_{min}$ (Wm$^{-1}$K$^{-1}$) |
|---|---|---|---|---|---|---|---|
| $CaPd_3Ti_4O_{12}$ | 5.94 | 7.93 | 3.99 | 4.48 | 608 | 2385.1 ± 300 | 1.30 |
| $CaPd_3V_4O_{12}$ | 6.34 | 7.64 | 3.89 | 4.36 | 604 | 2940.7 ± 300 | 1.31 |

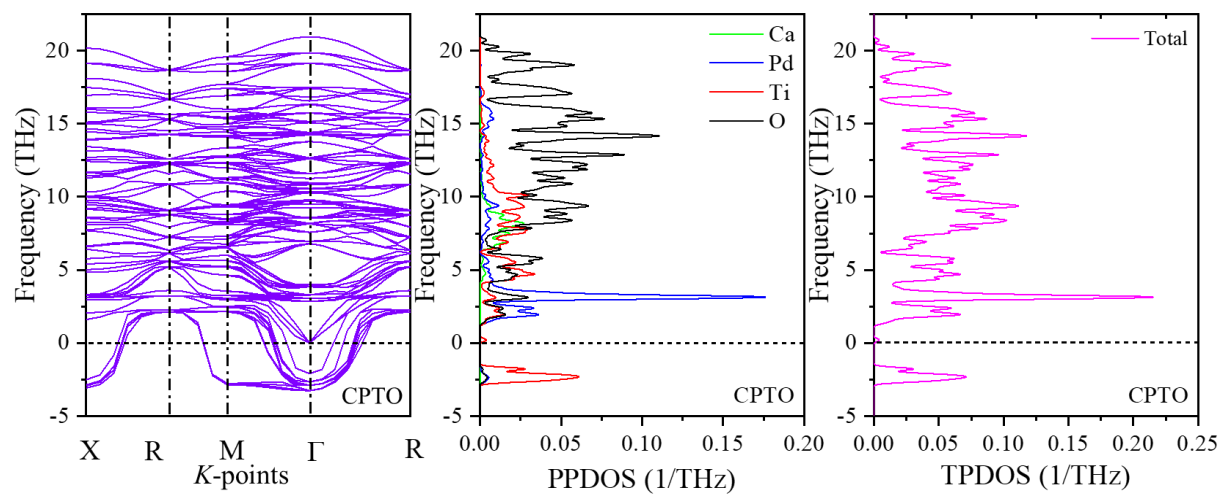

**Figure 9**. Computed phonon spectra/dispersion (left), total phonon density of states TPDOS (middle), and partial phonon density of states (right) of CPTO compound. The dotted line in the Figure indicates zero phonon frequency.

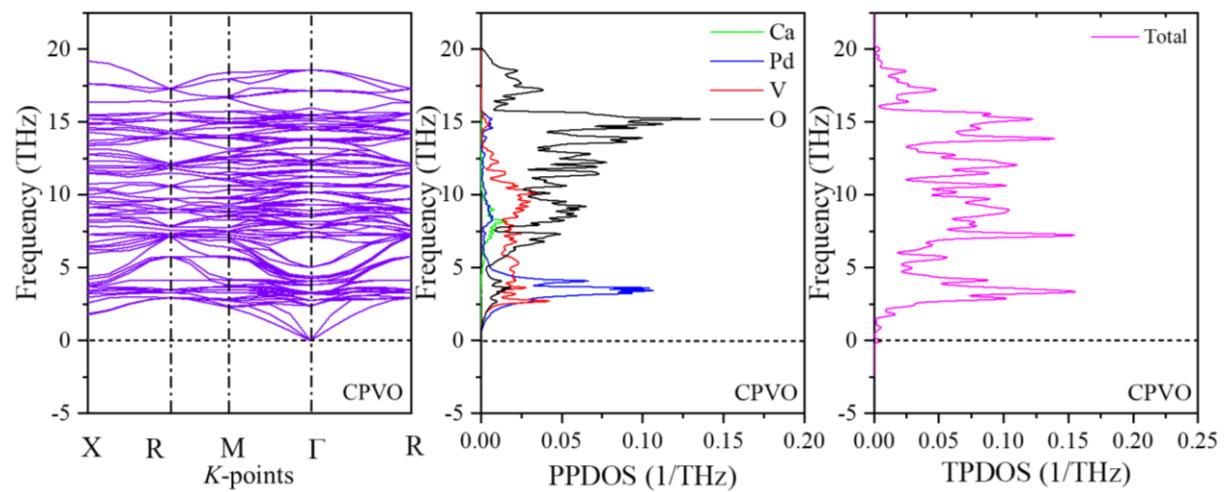

**Figure 10**. Computed phonon spectra/dispersion (left), total phonon density of states TPDOS (middle), and partial phonon density of states (right) of CPVO compound. The dotted line in the Figure indicates zero phonon frequency.

**Figures 9** and **10** show the predicted phonon dispersion curve (PDC) and density of phonon states for CPTO and CPVO along the main direction of the Brillouin zone (BZ) of the structure. The CP*B*O quadruple perovskite is a complex system that has twenty-two atoms; thus, the whole number of phonon modes is sixty-six (three times of total atoms). Interestingly, the PDCs of CP*B*O splits into two frequency segments like the low-frequency region and high -the frequency region where Ca, Pd, and Ti/V atoms show dispersion in the low-frequency range, whereas the O atom exhibits dispersion in the high-frequency region (**Figures 9** and **10**, right curves). The calculated phonon properties of the CPTO compound exhibit unstable dispersion in the aspect of phonon modes because several phonon dispersions are found below the zero-frequency (imaginary frequency) line at X, M, and Γ points along the M-Γ direction (**Figure 9**, left graph). Such type of unstable phonon modes in the material not only depend on the calculation method but also rather on structural and compositional geometry that primarily originates from the structural distortions, bond lengths (BVS) as well as bond angles, and oxygen octahedral rotation amid ions/atoms of materials [68,80]. This indicates that the slightly unstable phonon dispersion in CPTO might be originated from the slightly lower symmetry points, structural distortion (bond lengths and angles), and/or octahedral/tetrahedral rotation/distortion [68,80] due to the existence of Ti in the structure. Therefore, the slightly unstable phonon dispersion spectra for a particular element do not imply the dynamical instability of a compound, it further indicates the material might have a chance to undergo phase transition (new symmetry) after a finite applied temperature/field, as the corresponding imaginary phonon modes are observed in the PDC. Interestingly, the calculated phonon property of CPVO exhibits stable phonon dispersion at Γ-noddle point in contrast to the slight unstable phonon dispersion of CPTO. This stability in phonon dispersion for CPVO might be observed for the higher structural symmetry points owing to the presence of V in the place of Ti. Very recently, similar stable phonon dispersions like CPVO have also been investigated for several materials [81–83]. Furthermore, we calculated the phonon partial DOS (PPHDOS) of CPTO and CPVO to realize the atomic contribution in PDC, as shown in **Figures 9** and **10** (right curves). It is seen that the peaks of PPHDOS occur for the correlative procumment of bands, where the upwards and downwards bands alleviate the peak head in the TPDOS (middle curves). In **Figures 9 and 10** (right curves), we observe that in the low-frequency section up to 5 THz, the Pd atom has the highest phonon DOS, while Ti and V have a moderate contribution with trivial vibrational DOS of Ca and O atoms. In the frequency range of 5 to 10 THz, the primal contribution originates from the phonon modes of Ca, O, and Ti/V atoms. On the other hand, in the high-frequency region (≥ 10 THz) O atom is potently governed by the vibrational modes in PDC [68,80].

### 3.6 Thermoelectric transport properties of CaPd$_3$Ti$_4$O$_{12}$

The flat and narrow band gap semiconductors are potential candidates for thermoelectric device applications. The electronic band structure of the studied compounds reveals that CPVO is metallic whereas, CPTO is semiconducting. The narrow band gap of CPTO inspired us to study its thermoelectric transport properties. Recently, the thermoelectric properties of numerous semiconductors have been investigated based on DFT calculations [61,62,84–86]. The temperature-dependent thermoelectric transport features of CPTO are estimated using GGA-PBE and TB-mBJ approximations are presented in **Figure 11**. A larger Seebeck coefficient (S) and electrical conductivity (σ/τ) are essential in a possible thermoelectric material for a superior power factor, though their coexistence is unusual. The calculated S drops as the temperature rises for both materials. For GGA-PBE potential the calculated S is lower than that obtained using TB-mBJ potential. This is to be expected, given that the GGA-PBE potential undervalues the band gap. The obtained S at 300 K using GGA-PBE and TB-mBJ functional are 204 and 233 mV/K, respectively (**Figure 11(a)**). **Figure 11(b)** and **Figure 11(d)** depict the temperature dependence of electrical conductivity (σ/τ) and electronic thermal conductivity ($\kappa_e$/τ) and we notice that both are increasing with temperature. The increasing temperature enhances carrier concentration, and hence both the σ/τ and $\kappa_e$/τ increase. This behavior also indicates that CPTO is a semiconducting perovskite. The maximum power factor (S$^2$σ/τ) is 11.9 mWcm$^{-1}$K$^{-2}$ (with τ =10$^{-14}$ s) using TB-mBJ potential obtained at 800 K as shown in **Figure 11(c)** is slightly higher than that of SnSe, a promising thermoelectric material [87]. The dimensionless figure of merit (ZT) is directly concerned with both electronic thermal conductivity as well as lattice thermal conductivity. **Figure 11(e)** shows the ZT behavior in terms of electronic thermal conductivity and it is observed that ZT ~0.8 reaches a maximum in the temperature range of 800 to 1100 K. However, in this investigation we did not calculate lattice thermal conductivity which is directly related to lattice vibration (phonon scattering). In general, at low temperatures due to low phonon scattering lattice, thermal conductivity is much higher compared to electronic thermal conductivity but at a higher temperature it decreases rapidly and electronic thermal conductivity is dominating in semiconducting materials. Therefore, CPTO is a promising material for thermoelectric energy conversion and it might be encouraged experimentalists to measure thermoelectric properties for practical uses.

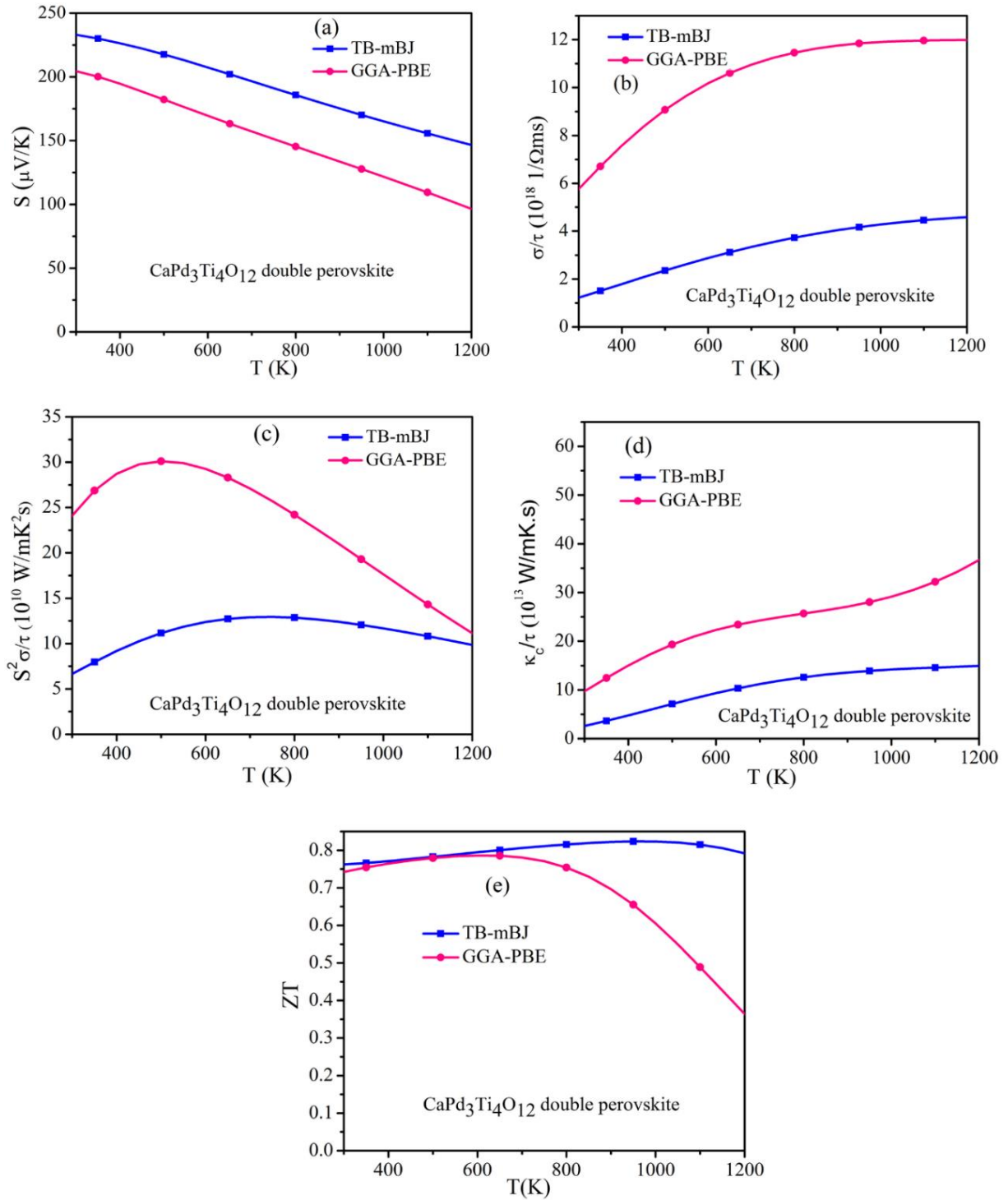

**Figure 11.** Thermoelectric transport behaviors as a function of the temperature of $CaPd_3Ti_4O_{12}$ double perovskite using both GGA-PBE and TB-mBJ potentials (a) Seebeck coefficient, (b) Electrical conductivity, (c) Power factor, (d) Electronic thermal conductivity, and (e) Dimensionless figure of merit.

## 4   Conclusions

DFT methods were used to investigate the structural, mechanical, and elastic properties, as well as electronic, optical, thermodynamic, phonon, and thermoelectric characters of the synthesized quadruple cubic perovskites $CaPd_3Ti_4O_{12}$ and $CaPd_3V_4O_{12}$. The calculated structural parameters of these perovskites logically corroborate the reported experimental data that clarify the authenticity of the present computations. The calculated elastic constants also satisfy the mechanical (Born stability criteria) and structural stability conditions for both compounds. The computed polycrystalline elastic constants and universal anisotropy index imply moderate hardness and elastically anisotropy of CP$B$O ($B$ = Ti, V). Moreover, Poisson's ratio, Pugh's ratio, and Cauchy pressure reveal their ductile nature. The calculated band structures of $CaPd_3Ti_4O_{12}$ exhibit a direct band gap of ~0.88 eV and 0.46 eV using GGA-mBJ and GGA-PBE potentials, respectively. On the other hand, the band structure of $CaPd_3V_4O_{12}$ reflects its metallic nature. Based on the computed partial density of states, Pd-4$d$ and O-2$p$ orbital electrons for CPTO and Pd-4$d$ and V-3$d$-O-2$p$ for CPVO exhibit substantial hybridization. The bonding properties reveal the metallic and coexistence of ionic and covalent bonds of relevant elements that have been seen from charge density maps and bonding population analysis. Furthermore, the multi−band feature of $CaPd_3V_4O_{12}$ is noticed from its Fermi surface topology. The static dielectric constants of 11.07 ($n(0)$ = 3.23)) and 5.78 ($n(0)$ = 9.78)) were observed for $CaPd_3Ti_4O_{12}$ and $CaPd_3V_4O_{12}$, respectively. Moreover, the high values of reflectivity of these compounds suggest that the materials have a promising coating application to attenuate solar heating. Very similar trends in photoconductivity and absorption spectra are visualized for both quadruple perovskites and their conductivity increases with absorbing photons. However, the calculated optical properties display slight anisotropy with respect to the used photon energy. Moreover, various thermodynamic properties have been calculated, which confirm the dynamical stability of both materials. In the midst of them, Debye temperature and melting temperature exhibit reasonably high values, where a larger Debye temperature comprises a better phonon thermal conductivity and vice versa. However, the phonon dispersion and phonon density of states of CPVO manifest a stable nature in contrast to the slightly unstable phonon properties of CPTO. Among the thermoelectric properties, the Seebeck coefficient gradually reduces at increasing temperature. Electronic conductivity and electronic thermal conductivity are both proportional to carrier concentration and increase with rising temperature. The estimated power factor increases gradually with the increase in temperature and then decreases again after 800 K. The maximum power factor ~11.9 mWcm$^{-1}$K$^{-2}$ with t = 10$^{-14}$ s was found at 800 K. The predicted thermoelectric figure of merit (ZT = 0.8, by using electronic thermal conductivity) at 800 K is very close to unity, which makes the studied material auspicious for thermoelectric device application. However, the reduction of thermal conductivity or enhancement of power factor can be achieved by nano-structuring or doping for practical use in converting waste heat to electricity in the CPTO system.

**Author's contributions**

**M.H.K. Rubel:** Conceptualization, supervision, writing manuscript and review editing; **M. A. Hossain:** Calculation of thermoelectric properties, data analysis, writing manuscript draft, review editing; **M.K. Hossain:** Validation, Writing – review & editing; **K.M. Hossain:** Calculations and analysis, draft writing, review editing; **A.A. Khatun**: Calculations and analysis; **M.M. Rahaman:** Conceptualization, formal analysis, review editing**; M.F. Rahaman: R**eview editing; **M. M. Hossain:** Calculations, Formal analysis, review editing; **J. Hossain:** Review editing.

**Data availability**

All data are used to evaluate the conclusion of this study are presented in the manuscript. Additional data can be available from the corresponding authors upon reasonable request.

**Conflict of Interests**

The authors declare no conflict of interest.

**Acknowledgments**

This research work was supported by a grant (No. 106/5/52/R.U./Eng.) from the Faculty of Engineering, University of Rajshahi, Rajshahi 6205, Bangladesh.

**References**

[1] Martínez-Lope MJ, Alonso JA, Casais MT. Synthesis, Crystal and Magnetic Structure of the New Double Perovskite $Ba_2MnMoO_6$. Zeitschrift Für Naturforsch B 2003;58:571–6. doi:10.1515/znb-2003-0613.

[2] Hong K-P, Choi Y-H, Kwon Y-U, Jung D-Y, Lee J-S, Shim H-S, et al. Atomic and Magnetic Long-Range Orderings in $BaLaMRuO_6$ (M=Mg and Zn). J Solid State Chem 2000;150:383–90. doi:10.1006/jssc.1999.8611.

[3] Ganguli AK, Grover V, Thirumal M. New double perovskites having low dielectric loss: LaBaZnTaO$_6$, LaSrZnNbO$_6$, and Ba$_2$Zn$_{0.5}$Ti$_{0.5}$TaO$_6$. Mater Res Bull 2001;36:1967–75. doi:10.1016/S0025-5408(01)00675-4.

[4] Seymour ID, Chroneos A, Kilner JA, Grimes RW. Defect processes in orthorhombic LnBaCo$_2$O$_{5.5}$ double perovskites. Phys Chem Chem Phys 2011;13:15305. doi:10.1039/c1cp21471c.

[5] Tarancón A, Skinner SJ, Chater RJ, Hernández-Ramírez F, Kilner JA. Layered perovskites as promising cathodes for intermediate temperature solid oxide fuel cells. J Mater Chem 2007;17:3175. doi:10.1039/b704320a.

[6] Parfitt D, Chroneos A, Tarancón A, Kilner JA. Oxygen ion diffusion in cation ordered/disordered GdBaCo$_2$O$_{5+\delta}$. J Mater Chem 2011;21:2183–6. doi:10.1039/C0JM02924F.

[7] Hossain MK, Biswas MC, Chanda RK, Rubel MHK, Khan MI, Hashizume K. A review on experimental and theoretical studies of perovskite barium zirconate proton conductors. Emergent Mater 2021;4:999–1027. doi:10.1007/s42247-021-00230-5.

[8] Hossain MK, Tamura H, Hashizume K. Visualization of hydrogen isotope distribution in yttrium and cobalt doped barium zirconates. J Nucl Mater 2020;538:152207. doi:10.1016/j.jnucmat.2020.152207.

[9] Hossain MK, Chanda R, El-Denglawey A, Emrose T, Rahman MT, Biswas MC, et al. Recent progress in barium zirconate proton conductors for electrochemical hydrogen device applications: A review. Ceram Int 2021;47:23725–48. doi:10.1016/j.ceramint.2021.05.167.

[10] Hossain MK, Iwasa T, Hashizume K. Hydrogen isotope dissolution and release behavior in Y-doped BaCeO3. J Am Ceram Soc 2021;104:6508–20. doi:10.1111/jace.18035.

[11] Azad A., Eriksson S-G, Ivanov S., Mathieu R, Svedlindh P, Eriksen J, et al. Synthesis, structural and magnetic characterisation of the double perovskite A$_2$MnMoO$_6$ (A=Ba, Sr). J Alloys Compd 2004;364:77–82. doi:10.1016/S0925-8388(03)00611-X.

[12] Liu XJ, Huang QJ, Zhang SY, Luo AH, Zhao CX. Thermal diffusivity of double perovskite Sr$_2$MMoO$_6$ (M=Fe, Mn and Co) and La doping effects studied by mirage effect. J Phys Chem Solids 2004;65:1247–51. doi:10.1016/j.jpcs.2004.01.017.

[13] Momma K, Izumi F. VESTA : a three-dimensional visualization system for electronic and structural analysis. J Appl Crystallogr 2008;41:653–8. doi:10.1107/S0021889808012016.

[14] Shiro K, Yamada I, Ikeda N, Ohgushi K, Mizumaki M, Takahashi R, et al. Pd$^{2+}$ -Incorporated Perovskite CaPd$_3$B$_4$O$_{12}$ ( B = Ti, V). Inorg Chem 2013;52:1604–9. doi:10.1021/ic3025155.

[15] Subramanian MA, Li D, Duan N, Reisner BA, Sleight AW. High Dielectric Constant in ACu$_3$Ti$_4$O$_{12}$ and ACu$_3$Ti$_3$FeO$_{12}$ Phases. J Solid State Chem 2000;151:323–5. doi:10.1006/jssc.2000.8703.

[16] Shimakawa Y. A-Site-Ordered Perovskites with Intriguing Physical Properties. Inorg Chem 2008;47:8562–70. doi:10.1021/ic800696u.

[17] Rubel MHK, Miura A, Takei T, Kumada N, Mozahar Ali M, Nagao M, et al. Superconducting Double Perovskite Bismuth Oxide Prepared by a Low-Temperature Hydrothermal Reaction. Angew Chemie 2014;126:3673–7. doi:10.1002/ange.201400607.

[18] Jiang H, Kumada N, Yonesaki Y, Takei T, Kinomura N, Yashima M, et al. Hydrothermal Synthesis of a New Double Perovskite-Type Bismuthate, (Ba$_{0.75}$K$_{0.14}$H$_{0.11}$)BiO$_3$·nH$_2$O. Jpn J Appl Phys 2009;48:010216. doi:10.1143/JJAP.48.010216.

[19] Rubel MHK, Takei T, Kumada N, Ali MM, Miura A, Tadanaga K, et al. Hydrothermal Synthesis, Crystal Structure, and Superconductivity of a Double-Perovskite Bi Oxide. Chem Mater 2016;28:459–65. doi:10.1021/acs.chemmater.5b02386.

[20] Vasil'ev AN, Volkova OS. New functional materials AC$_3$B$_4$O$_{12}$ (Review). Low Temp Phys 2007;33:895–914. doi:10.1063/1.2747047.

[21] Ali MS, Aftabuzzaman M, Roknuzzaman M, Rayhan MA, Parvin F, Ali MM, et al. New superconductor (Na$_{0.25}$K$_{0.45}$)Ba$_3$Bi$_4$O$_{12}$: A first-principles study. Phys C Supercond Its Appl 2014;506:53–8. doi:10.1016/j.physc.2014.08.010.

[22] Rubel MHK, Mozahar Ali M, Ali MS, Parvin R, Rahaman MM, Hossain KM, et al. First−principles study: Structural, mechanical, electronic and thermodynamic properties of simple−cubic−perovskite $(Ba_{0.62}K_{0.38})(Bi_{0.92}Mg_{0.08})O_3$. Solid State Commun 2019;288:22–7. doi:10.1016/j.ssc.2018.11.008.

[23] Rubel MHK, Hadi MA, Rahaman MM, Ali MS, Aftabuzzaman M, Parvin R, et al. Density functional theory study of a new Bi-based $(K_{1.00})(Ba_{1.00})_3(Bi_{0.89}Na_{0.11})_4O_{12}$ double perovskite superconductor. Comput Mater Sci 2017;138:160–5. doi:10.1016/j.commatsci.2017.06.030.

[24] Rubel M, Mitro SK, Hossain KMM, Rahaman MM, Hossain MK, Hossain J, et al. A Comprehensive First Principles Calculations On $(Ba_{0.82}K_{0.18})(Bi_{0.53}Pb_{0.47})O_3$ Single-Cubic-Perovskite Superconductor. SSRN Electron J 2021. doi:10.2139/ssrn.3984132.

[25] Clark SJ, Segall MD, Pickard CJ, Hasnip PJ, Probert MIJ, Refson K, et al. First principles methods using CASTEP. Zeitschrift Für Krist - Cryst Mater 2005;220:567–70. doi:10.1524/zkri.220.5.567.65075.

[26] Radja K, Farah BL, Ibrahim A, Lamia D, Fatima I, Nabil B, et al. Investigation of structural, magneto-electronic, elastic, mechanical and thermoelectric properties of novel lead-free halide double perovskite $Cs_2AgFeCl_6$: First-principles calcuations. J Phys Chem Solids 2022;167:110795. doi:10.1016/j.jpcs.2022.110795.

[27] Fadila B, Ameri M, Bensaid D, Noureddine M, Ameri I, Mesbah S, et al. Structural, magnetic, electronic and mechanical properties of full-Heusler alloys $Co_2YAl$ (Y = Fe, Ti): First principles calculations with different exchange-correlation potentials. J Magn Magn Mater 2018;448:208–20. doi:10.1016/j.jmmm.2017.06.048.

[28] Khireddine A, Bouhemadou A, Alnujaim S, Guechi N, Bin-Omran S, Al-Douri Y, et al. First-principles predictions of the structural, electronic, optical and elastic properties of the zintl-phases $AE_3GaAs_3$ (AE = Sr, Ba). Solid State Sci 2021;114:106563. doi:10.1016/j.solidstatesciences.2021.106563.

[29] Mentefa A, Boufadi FZ, Ameri M, Gaid F, Bellagoun L, Odeh AA, et al. First-Principles Calculations to Investigate Structural, Electronic, Elastic, Magnetic, and Thermodynamic Properties of Full-Heusler $Rh_2MnZ$ (Z = Zr, Hf). J Supercond Nov Magn 2021;34:269–83. doi:10.1007/s10948-020-05741-6.

[30] Ayad M, Belkharroubi F, Boufadi FZ, Khorsi M, Zoubir MK, Ameri M, et al. First-principles calculations to investigate magnetic and thermodynamic properties of new multifunctional full-Heusler alloy $Co_2TaGa$. Indian J Phys 2020;94:767–77. doi:10.1007/s12648-019-01518-3.

[31] Hadji S, Bouhemadou A, Haddadi K, Cherrad D, Khenata R, Bin-Omran S, et al. Elastic, electronic, optical and thermodynamic properties of $Ba_3Ca_2Si_2N_6$ semiconductor: First-principles predictions. Phys B Condens Matter 2020;589:412213. doi:10.1016/j.physb.2020.412213.

[32] Al-Douri Y, Ameri M, Bouhemadou A, Batoo KM. First-Principles Calculations to Investigate the Refractive Index and Optical Dielectric Constant of $Na_3SbX_4$ ( X = S, Se) Ternary Chalcogenides. Phys Status Solidi 2019;256:1900131. doi:10.1002/pssb.201900131.

[33] Boudiaf K, Bouhemadou A, Al-Douri Y, Khenata R, Bin-Omran S, Guechi N. Electronic and thermoelectric properties of the layered BaFAgCh ( Ch = S, Se and Te): First-principles study. J Alloys Compd 2018;759:32–43. doi:10.1016/j.jallcom.2018.05.142.

[34] Belkilali W, Belkharroubi F, Ameri M, Ramdani N, Boudahri F, Khelfaoui F, et al. Theoretical investigations of structural, mechanical, electronic and optical properties of NaScSi alloy. Emergent Mater 2021;4:1465–77. https://doi.org/10.1007/s42247-021-00221-6.

[35] Ameri M, Bennar F, Amel S, Ameri I, Al-Douri Y, Varshney D. Structural, elastic, thermodynamic and electronic properties of LuX (X = N, Bi and Sb) compounds: first principles calculations. Phase Transitions 2016;89:1236–52. https://doi.org/10.1080/01411594.2016.1162791.

[36] Sugahara T, Ohtaki M, Souma T. Thermoelectric properties of double-perovskite oxide $Sr_{2-x}M_xFeMoO_6$ (M = Ba, La). J Ceram Soc Japan 2008;116:1278–82. doi:10.2109/jcersj2.116.1278.

[37] Sugahara T, Araki T, Ohtaki M, Suganuma K. Structure and thermoelectric properties of double-perovskite oxides: $Sr_{2-x}K_xFeMoO_6$. J Ceram Soc Japan 2012;120:211–6. doi:10.2109/jcersj2.120.211.

[38] Sugahara T, Ohtaki M. Structural and semiconductor-to-metal transitions of double-perovskite cobalt oxide $Sr_{2-x}La_xCoTiO_{6-\delta}$ with enhanced thermoelectric capability. Appl Phys Lett 2011;99:062107. doi:10.1063/1.3623476.

[39] Ali MA, Anwar Hossain M, Rayhan MA, Hossain MM, Uddin MM, Roknuzzaman M, et al. First-principles study of elastic, electronic, optical and thermoelectric properties of newly synthesized $K_2Cu_2GeS_4$ chalcogenide. J Alloys Compd 2019;781:37–46. doi:10.1016/j.jallcom.2018.12.035.

[40] Segall MD, Lindan PJD, Probert MJ, Pickard CJ, Hasnip PJ, Clark SJ, et al. First-principles simulation: ideas, illustrations and the CASTEP code. J Phys Condens Matter 2002;14:2717–44. doi:10.1088/0953-8984/14/11/301.

[41] Blaha P, Schwarz K, Madsen GKH, Kvasnicka D, Luitz J, Laskowski R, et al. WIEN2k: An Augmented PlaneWave Plus Local Orbitals Program for Calculating Crystal Properties, User's Guide, WIEN2k 21.1. Vienna, Austria: Vienna University of Technology; 2021.

[42] Rahaman MM, Hossain KM, Rubel MHK, Islam AKMA, Kojima S. Alkaline-Earth Metal Effects on Physical Properties of Ferromagnetic $AVO_3$ (A = Ba, Sr, Ca, and Mg): Density Functional Theory Insights. ACS Omega 2022;7:20914–26. doi:10.1021/acsomega.2c01630.

[43] Payne MC, Teter MP, Allan DC, Arias TA, Joannopoulos JD. Iterative minimization techniques for ab initio total-energy calculations: molecular dynamics and conjugate gradients. Rev Mod Phys 1992;64:1045–97. doi:10.1103/RevModPhys.64.1045.

[44] Tran F, Blaha P. Accurate Band Gaps of Semiconductors and Insulators with a Semilocal Exchange-Correlation Potential. Phys Rev Lett 2009;102:226401. doi:10.1103/PhysRevLett.102.226401.

[45] Perdew JP, Ruzsinszky A, Csonka GI, Vydrov OA, Scuseria GE, Constantin LA, et al. Restoring the Density-Gradient Expansion for Exchange in Solids and Surfaces. Phys Rev Lett 2008;100:136406. doi:10.1103/PhysRevLett.100.136406.

[46] Vanderbilt D. Soft self-consistent pseudopotentials in a generalized eigenvalue formalism. Phys Rev B 1990;41:7892–5. doi:10.1103/PhysRevB.41.7892.

[47] Monkhorst HJ, Pack JD. Special points for Brillouin-zone integrations. Phys Rev B 1976;13:5188–92. doi:10.1103/PhysRevB.13.5188.

[48] Fischer TH, Almlof J. General methods for geometry and wave function optimization. J Phys Chem 1992;96:9768–74. doi:10.1021/j100203a036.

[49] Liu SS-Y, Zhang S, Liu SS-Y, Li D-J, Li Y, Wang S. Phase stability, mechanical properties and melting points of high-entropy quaternary metal carbides from first-principles. J Eur Ceram Soc 2021;41:6267–74. doi:10.1016/j.jeurceramsoc.2021.05.022.

[50] Jamal M. 2DRoptimize package is provided by M. Jamal as part of the commercial code WIEN2K 2014.

[51] Madsen GKH, Singh DJ. BoltzTraP. A code for calculating band-structure dependent quantities. Comput Phys Commun 2006;175:67–71. doi:10.1016/j.cpc.2006.03.007.

[52] Perdew JP, Burke K, Ernzerhof M. Generalized Gradient Approximation Made Simple. Phys Rev Lett 1996;77:3865–8. doi:10.1103/PhysRevLett.77.3865.

[53] Bouhemadou A. First-principles study of structural, electronic and elastic properties of $Nb_4AlC_3$. Brazilian J Phys 2010;40:52–7. doi:10.1590/S0103-97332010000100009.

[54] Chen X-Q, Niu H, Li D, Li Y. Modeling hardness of polycrystalline materials and bulk metallic glasses. Intermetallics 2011;19:1275–81. doi:10.1016/j.intermet.2011.03.026.

[55] Zhou P, Gong HR. Phase stability, mechanical property, and electronic structure of an Mg–Ca system. J Mech Behav Biomed Mater 2012;8:154–64. doi:10.1016/j.jmbbm.2011.12.001.

[56] Bouhemadou A, Khenata R, Kharoubi M, Seddik T, Reshak AH, Al-Douri Y. FP-APW+lo calculations of the elastic properties in zinc-blende III-P compounds under pressure effects. Comput Mater Sci 2009;45:474–9. doi:10.1016/j.commatsci.2008.11.013.

[57] Pettifor DG. Theoretical predictions of structure and related properties of intermetallics. Mater Sci Technol 1992;8:345–9. doi:10.1179/mst.1992.8.4.345.

[58] Ranganathan SI, Ostoja-Starzewski M. Universal Elastic Anisotropy Index. Phys Rev Lett 2008;101:055504. doi:10.1103/PhysRevLett.101.055504.

[59] Lu G. The Peierls—Nabarro Model of Dislocations: A Venerable Theory and its Current Development.

Handb. Mater. Model., Dordrecht: Springer Netherlands; 2005, p. 793–811. doi:10.1007/978-1-4020-3286-8_41.

[60] Music D, Schneider JM. Elastic properties of $Sr_{n+1}Ti_nO_{3n+1}$ phases ( n = 1–3, \infty ). J Phys Condens Matter 2008;20:055224. doi:10.1088/0953-8984/20/5/055224.

[61] Guechi N, Bouhemadou A, Bin-Omran S, Bourzami A, Louail L. Elastic, Optoelectronic and Thermoelectric Properties of the Lead-Free Halide Semiconductors $Cs_2AgBiX_6$ (X = Cl, Br): Ab Initio Investigation. J Electron Mater 2018;47:1533–45. doi:10.1007/s11664-017-5962-2.

[62] Irfan M, Azam S, Hussain S, Khan SA, Sohail M, Ahmad M, et al. Enhanced thermoelectric properties of $ASbO_3$ due to decreased band gap through modified becke johnson potential scheme. J Phys Chem Solids 2018;119:85–93. doi:10.1016/j.jpcs.2018.03.010.

[63] Hu W-C, Liu Y, Li D-J, Zeng X-Q, Xu C-S. First-principles study of structural and electronic properties of C14-type Laves phase Al2Zr and Al2Hf. Comput Mater Sci 2014;83:27–34. doi:10.1016/j.commatsci.2013.10.029.

[64] Okamoto Y, Inohara T, Yamakawa Y, Yamakage A, Takenaka K. Superconductivity in the Hexagonal Ternary Phosphide ScIrP. J Phys Soc Japan 2016;85:013704. doi:10.7566/JPSJ.85.013704.

[65] Inohara T, Okamoto Y, Yamakawa Y, Takenaka K. Synthesis and Superconducting Properties of a Hexagonal Phosphide ScRhP. J Phys Soc Japan 2016;85:094706. doi:10.7566/JPSJ.85.094706.

[66] Rahaman MM, Rubel MHK, Rashid MA, Alam MA, Hossain KM, Hossain MI, et al. Mechanical, electronic, optical, and thermodynamic properties of orthorhonmbic $LiCuBiO_4$ crystal: a first–priciples study. J Mater Res Technol 2019;8:3783–94. doi:10.1016/j.jmrt.2019.06.039.

[67] Singh RP. First principle study of structural, electronic and thermodynamic behavior of ternary intermetallic compound: CeMgTl. J Magnes Alloy 2014;2:349–56. doi:10.1016/j.jma.2014.10.004.

[68] Lebedev AI. Ab initio calculations of phonon spectra in $ATiO_3$ perovskite crystals (A = Ca, Sr, Ba, Ra, Cd, Zn, Mg, Ge, Sn, Pb). Phys Solid State 2009;51:362–72. doi:10.1134/S1063783409020279.

[69] Mondal BK, Newaz MA, Rashid MA, Hossain KM, Mostaque SK, Rahman MF, et al. Electronic Structure of $In_{3-x}Se_4$ Electron Transport Layer for Chalcogenide/p-Si Heterojunction Solar Cells. ACS Omega 2019;4:17762–72. doi:10.1021/acsomega.9b02210.

[70] Yang J, Yang L, Long J. Theoretical investigation of the electronic structure, optical, elastic, hardness and thermodynamics properties of jadeite. Mater Sci Semicond Process 2015;31:509–16. doi:10.1016/j.mssp.2014.12.039.

[71] Okoye CMI. Optical properties of the antiperovskite superconductor $MgCNi_3$. J Phys Condens Matter 2003;15:833–41. doi:10.1088/0953-8984/15/6/310.

[72] Sun J, Zhou X-F, Fan Y-X, Chen J, Wang H-T, Guo X, et al. First-principles study of electronic structure and optical properties of heterodiamond BC2N. Phys Rev B 2006;73:045108. doi:10.1103/PhysRevB.73.045108.

[73] Parvin F, Hossain MA, Ali MS, Islam AKMA. Mechanical, electronic, optical, thermodynamic properties and superconductivity of ScGa. Phys B Condens Matter 2015;457:320–5. doi:10.1016/j.physb.2014.10.007.

[74] Fox M. Optical Properties of Solids. New York, USA: Academic Press; n.d.

[75] Mulliken RS. Electronic Population Analysis on LCAO–MO Molecular Wave Functions. II. Overlap Populations, Bond Orders, and Covalent Bond Energies. J Chem Phys 1955;23:1841–6. doi:10.1063/1.1740589.

[76] Segall MD, Shah R, Pickard CJ, Payne MC. Population analysis of plane-wave electronic structure calculations of bulk materials. Phys Rev B 1996;54:16317–20. doi:10.1103/PhysRevB.54.16317.

[77] Anderson OL. A simplified method for calculating the debye temperature from elastic constants. J Phys Chem Solids 1963;24:909–17. doi:10.1016/0022-3697(63)90067-2.

[78] Alouani M, Albers RC, Methfessel M. Calculated elastic constants and structural properties of Mo and MoSi2. Phys Rev B 1991;43:6500–9. doi:10.1103/PhysRevB.43.6500.

[79] Clarke DR. Materials selection guidelines for low thermal conductivity thermal barrier coatings. Surf

Coatings Technol 2003;163–164:67–74. doi:10.1016/S0257-8972(02)00593-5.

[80] Hossain KM, Rubel MHK, Rahaman MM, Hossain MM, Hossain MI, Khatun AA, et al. A comparative theoretical study on physical properties of synthesized $AVO_3$ (A = Ba, Sr, Ca, Pb) perovskites 2019.

[81] Ding G, Zhou F, Zhang Z, Yu Z-M, Wang X. Charge-two Weyl phonons with type-III dispersion. Phys Rev B 2022;105:134303. doi:10.1103/PhysRevB.105.134303.

[82] Yang T, Xie C, Chen H, Wang X, Zhang G. Phononic nodal points with quadratic dispersion and multifold degeneracy in the cubic compound $Ta_3Sn$. Phys Rev B 2022;105:094310. doi:10.1103/PhysRevB.105.094310.

[83] Haring-Kaye RA, Palombi F, Döring J, Tabor SL, Abromeit B, Lubna R, et al. Onset of band structure in Ga70. Phys Rev C 2022;105:054307. doi:10.1103/PhysRevC.105.054307.

[84] Haque E, Hossain MA. Origin of ultra-low lattice thermal conductivity in $Cs_2BiAgX_6$ (X=Cl, Br) and its impact on thermoelectric performance. J Alloys Compd 2018;748:63–72. doi:10.1016/j.jallcom.2018.03.137.

[85] Rahman MT, Haque E, Hossain MA. Elastic, electronic and thermoelectric properties of $Sr_3MN$ (M = Sb, Bi) under pressure. J Alloys Compd 2019;783:593–600. doi:10.1016/j.jallcom.2018.12.339.

[86] Haque E, Hossain MA. First-principles study of elastic, electronic, thermodynamic, and thermoelectric transport properties of TaCoSn. Results Phys 2018;10:458–65. doi:10.1016/j.rinp.2018.06.053.

[87] Zhao L-D, Lo S-H, Zhang Y, Sun H, Tan G, Uher C, et al. Ultralow thermal conductivity and high thermoelectric figure of merit in SnSe crystals. Nature 2014;508:373–7. doi:10.1038/nature13184.